\documentclass[11pt, letterpaper]{article}
\pdfoutput=1
\usepackage{amsmath,amssymb,amsfonts,amsthm,bbm}
\usepackage{epic,eepic,epsfig,longtable}
\usepackage{multirow,verbatim}
\usepackage{array}
\usepackage{graphicx}
\usepackage{caption}
\usepackage{epsfig}
\usepackage{setspace}
\usepackage{bbm}
\usepackage{CJK}
\usepackage{color}
\usepackage[utf8x]{inputenc} 
\usepackage[authoryear,round]{natbib}
\usepackage{romannum}
\makeatletter
\def\singlespace{\def\baselinestretch{1}\@normalsize}

\makeatletter
\def\singlespace{\def\baselinestretch{1}\@normalsize}

\numberwithin{equation}{section}
\renewcommand{\hat}{\widehat}

\renewcommand{\hat}{\widehat}

\def\ba{\bfm a}     
\def\bb{\bfm b}

\def\today{\ifcase\month\or
  January\or February\or March\or April\or May\or June\or
  July\or August\or September\or October\or November\or December\fi
  \space\number\day, \number\year}

\newdimen\biblioindent    \biblioindent=30pt

 at 8truept

\def\si{\sigma}
\newcommand{\beq}{\begin{equation}}
  \newcommand{\eeq}{\end{equation}}
\newcommand{\beqn}{\begin{eqnarray}}
  \newcommand{\eeqn}{\end{eqnarray}}
\newcommand{\beqnn}{\begin{eqnarray*}}
  \newcommand{\eeqnn}{\end{eqnarray*}}

\allowdisplaybreaks
\usepackage{verbatim}
\renewcommand{\baselinestretch}{1.66}
\newtheorem{lemma}{Lemma}
\newtheorem{assumption}{Assumption}
\newtheorem{definition}{Definition}
\newtheorem{theorem}{Theorem}
\newtheorem{proposition}{Proposition}
\newtheorem{remark}{Remark}
\newcounter{CondCounter}
\newcommand{\lonenorm}[1]{\left\lVert#1\right\rVert_{L_1}}
\newcommand{\ltwonorm}[1]{\left\lVert#1\right\rVert_{L_2}}
\newcommand{\lfournorm}[1]{\left\lVert#1\right\rVert_{L_4}}
\newcommand{\lpnorm}[1]{\left\lVert#1\right\rVert_{L_p}}
\newcommand{\maxnorm}[1]{\left\lVert#1\right\rVert_{max}}
\newcommand{\abs}[1]{\left|#1\right|}

\def \RVsig {\ltwonorm{RV_n-\isign}}

\def \sn {\sigma_n^2}
\def \sna {\sigma_{n-1}^2}
\def \snb {\sigma_{n-2}^2}
\def \isign {\int_{n-1}^n \sigma_t^2 dt}

\def \wi {\omega_i}
\def\gi {\gamma_i}
\def\betai {\beta_i}
\def\wa {\omega_1}
\def\ga {\gamma_1}
\def\ba {\beta_1}
\def\wb {\omega_2}
\def\gb {\gamma_2}
\def\bb {\beta_2}
\def \wah {\omega^h_{11}}
\def \wbh {\omega^h_{12}}
\def \wch {\omega^h_{21}}
\def \wdh {\omega^h_{22}}
\def \gah {\gamma^h_{11}}
\def \gbh {\gamma^h_{12}}
\def \gch {\gamma^h_{21}}
\def \gdh {\gamma^h_{22}}
\def \bah {\beta^h_{11}}
\def \bbh {\beta^h_{12}}
\def \bch {\beta^h_{21}}
\def \bdh {\beta^h_{22}}
\def \wuh {\omega^h_u}
\def \wlh {\omega^h_l}
\def \guh {\gamma^h_u}
\def \glh {\gamma^h_l}
\def \buh {\beta^h_u}
\def \blh {\beta^h_l}

\def \aw {\omega^w_{n}}
\def \ab {\beta^w_{n}}
\def \agij {\gamma^w_{n}}

\def \ago {\gamma^w_{0,n}}
\def \awp {\omega^w_{n-k+1}}
\def \abp {\beta^w_{n-k+1}}
\def \agp {\gamma^w_{n-k+1}}
\def \awop {\omega^w_{0,n-k+1}}
\def \abop {\beta^w_{0,n-k+1}}
\def \agop {\gamma^w_{0,n-k+1}}

\def \agq {\gamma^w_{n-l+1}}
\def \agoq {\gamma^w_{0,n-l+1}}

\def\woa {\omega_{0,1}}
\def\goa {\gamma_{0,1}}
\def\boa {\beta_{0,1}}
\def\wob {\omega_{0,2}}
\def\gob {\gamma_{0,2}}
\def\bob {\beta_{0,2}}
\def \woah {\omega^h_{0,11}}
\def \wobh {\omega^h_{0,12}}
\def \woch {\omega^h_{0,21}}
\def \wodh {\omega^h_{0,22}}
\def \goah {\gamma^h_{0,11}}
\def \gobh {\gamma^h_{0,12}}
\def \goch {\gamma^h_{0,21}}
\def \godh {\gamma^h_{0,22}}
\def \boah {\beta^h_{0,11}}
\def \bobh {\beta^h_{0,12}}
\def \boch {\beta^h_{0,21}}
\def \bodh {\beta^h_{0,22}}

\def \wahs {\omega^{h*}_{11}}
\def \wbhs {\omega^{h*}_{12}}
\def \wchs {\omega^{h*}_{21}}
\def \wdhs {\omega^{h*}_{22}}
\def \gahs {\gamma^{h*}_{11}}
\def \gbhs {\gamma^{h*}_{12}}
\def \gchs {\gamma^{h*}_{21}}
\def \gdhs {\gamma^{h*}_{22}}
\def \bahs {\beta^{h*}_{11}}
\def \bbhs {\beta^{h*}_{12}}
\def \bchs {\beta^{h*}_{21}}
\def \bdhs {\beta^{h*}_{22}}

\def \nint {\int_{n-1}^{n}}

\def \Zn {Z_n^2}
\def \Zna {Z_{n-1}^2}
\def \Znp {Z_{n-k}^2}

\def \Za {Z_{1,n}^2}
\def \Zb {Z_{2,n}^2}

\def \Zaa {Z_{1,n-1}^2}
\def \Zba {Z_{2,n-1}^2}
\def \yba {s_{n}}
\def \ybb {s_{n-1}}
\def \yaa {(1-s_{n})}
\def \yab {(1-s_{n-1})}
\def \yaya {s_{11,n}}
\def \yayb {s_{12,n}}
\def \ybya {s_{21,n}}
\def \ybyb {s_{22,n}}

\def \yayaoa {s_{11,1}}
\def \ybyboa {s_{12,1}}

\def \ybyboa {s_{22,1}}
\def \yayaob {s_{11,2}}

\def \ybybob {s_{22,2}}

\def \ha {h_{1,n}(\theta)}

\def \hb {h_{2,n}(\theta)}
\def \hn {h_n(\theta)}
\def \hon {h_n(\to)}
\def \hsn {h_n(\ts)}
\def \hna {h_{n-1}(\theta)}

\def \Hac {H_{c,1}(\theta)}
\def \Hab {H_{\beta,1}(\theta)}
\def \Hbc {H_{c,2}(\theta)}
\def \Hbb {H_{\beta,2}(\theta)}
\def \Hic {H_{c,i}(\theta)}
\def \Hib {H_{\beta,i}(\theta)}
\def \Hjc {H_{c,j}(\theta)}
\def \Hjb {H_{\beta,j}(\theta)}

\newcommand \aHc [1][n] {H^w_{c,#1}(\theta)}
\newcommand \aHb [1][n] {H^w_{\beta,#1}(\theta)}

\newcommand \dfunc [2] {{\partial #1 \over\partial #2}}
\def \dhj {{\partial\hn\over\partial\theta_j}}
\def \ddhjk {{\partial^2\hn\over\partial\theta_j\partial\theta_k}}
\def \dddhjkl {{\partial^3\hn\over\partial\theta_j\partial\theta_k\partial\theta_l}}
\def \dht {{\partial\hn\over\partial\theta}}
\def \dhtt {\left[{\partial\hn\over\partial\theta}\right]^T}
\def \dhot {{\partial\hon\over\partial\theta}}
\def \dhott {{\partial\hon\over\partial\theta^T}}
\def \dhst {{\partial\hsn\over\partial\theta}}

\def \ddhott {{\partial^2\hon\over\partial\theta\partial\theta^T}}

\def \th {\hat{\theta}}

\def \to {\theta_0}
\def \tolist {(\woa,\wob,\boa,\bob,\goa,\gob)}
\def \ts {\theta^*}

\def \bfilta {\Big|\mathcal{F}_{n-1}}
\def \sfilta {|\mathcal{F}_{n-1}}

\def \Lhnm {\hat{Q}_{N,M}(\theta)}

\def \Lhn {\widetilde{Q}_N(\theta)}

\def \lhn {\hat{q}_{N,M}(\theta)}

\def \Ln {Q_N(\theta)}

\def \phnm {\hat{S}_{N,M}(\theta)}
\def \phn {\widetilde{S}_N(\theta)}
\def \pn {S_N(\theta)}

\def \phonm {\hat{S}_{N,M}(\theta_0)}
\def \phon {\widetilde{S}_N(\theta_0)}
\def \pon {S_N(\theta_0)}
\def \dphonm {\triangledown\hat{S}_{N,M}(\theta_0)}
\def \dphon {\triangledown\widetilde{S}_N(\theta_0)}
\def \dpon {\triangledown{S}_N(\theta_0)}

\def \Gn {K_N(\theta)}
\def \Gnprime {K_N(\theta^{'})}

\def \sumn {\sum_{n=1}^{N}}
\def \sump {\sum_{k=1}^{n-1}}
\def \prodp {\prod_{k=1}^{n-1}}
\def \prodq {\prod_{l=1}^{k-1}}
\def \supn {\sup_{\ninN}}
\def \supt {\sup_{\tinT}}
\def \supB {\sup_{\theta\in B(\to)}}

\def \ninN {n\in\mathbb{N}}
\def \tinT {\theta\in\Theta}
\usepackage{textcomp}
\usepackage{tikz}
\usetikzlibrary{trees, shapes.geometric}
\usetikzlibrary{shapes.gates.logic.US,trees,positioning,arrows}
\usepackage{bbm}

\usepackage{epic,eepic,epsfig,longtable}
\usepackage{multirow,verbatim}
\usepackage{array}
\usepackage{graphicx}
\usepackage{caption}
\usepackage{epsfig}
\usepackage{bbm}
\usepackage{CJK}
\usepackage{color}
\usepackage{xcolor}

\usepackage{longtable}
\usepackage{booktabs}
\usepackage[utf8x]{inputenc} 
\usepackage[authoryear,round]{natbib}
\title{ State Heterogeneity Analysis of Financial Volatility\\ Using High-Frequency Financial Data}
\author{Dohyun Chun$^1$ and Donggyu Kim$^1$\footnote{corresponding author.
\newline
E-mail addresses: dohyun0323@kaist.ac.kr (D. Chun), donggyukim@kaist.ac.kr (D. Kim).} \\
\footnotesize{$^1$College of Business, Korea Advanced Institute of Science and Technology (KAIST), Seoul, Korea}\\
}
\date{\today}

\begin{document}
\pagenumbering{arabic}

%   \begin{center}
%       \vspace*{2cm}
%       \fontsize{14}{17}\selectfont
%
%       \textbf{State Heterogeneity Analysis of Financial Volatility\\ Using High-Frequency Financial Data}
%
%       
%
%       \vspace{2cm}
%       
%
%
%       Dohyun Chun and Donggyu Kim
%       
%       \vspace{2cm}
%       
%       \fontsize{12}{17}\selectfont
%       
%       College of Business, Korea Advanced Institute of Science and Technology (KAIST),\\ Seoul, Korea
%
%       \vspace{2.5cm}
%       
%    \end{center}
%    
%    \fontsize{12}{17}\selectfont
%    
%    \noindent
%    \textbf{Corrsponding Author:}\\
%       Donggyu Kim\\
%       Korea Advanced Institute of Science and Technology\\
%       85 Hegiro, Dongdaemoon-gu, Seoul 02455, Korea\\
%       E-mail: donggyukim@kaist.ac.kr\\
%       Tel: +82-2-958-3448\\
%
%\clearpage

 \maketitle

\begin{abstract}

\noindent
Recently, to account for low-frequency market dynamics, several volatility models, employing high-frequency financial data, have been developed. 
However, in financial markets, we often observe that financial volatility processes depend on economic states, so they have a state heterogeneous structure.
In this paper, to study state heterogeneous market dynamics based on high-frequency data, we introduce a novel volatility model based on a continuous It\^o diffusion process whose intraday instantaneous volatility process evolves depending on the exogenous state variable, as well as its integrated volatility.
We call it the state heterogeneous GARCH-It\^o (SG-It\^o) model.
We suggest a quasi-likelihood estimation procedure with the realized volatility proxy and establish its asymptotic behaviors. 
Moreover, to test the low-frequency state heterogeneity, we develop a Wald test-type hypothesis testing procedure. 
The results of empirical studies suggest the existence of leverage, investor attention, market illiquidity, stock market comovement, and post-holiday effect in S\&P 500 index volatility.

\end{abstract}



\section{Introduction} \label{Introduction}

 %is an essential state variable in the distribution of asset return and 
Volatility plays an important role in financial asset pricing, risk management, portfolio allocation, and managerial decision-making.
These interests have led many researchers to analyze financial volatility features such as time-varying heteroscedasticity, heavy tailness, and volatility clustering effect. 
To account for stylized market features, GARCH models \citep{bollerslev1986generalized, engle1982autoregressive} have been introduced. 
In financial markets, we often observe that volatility varies with economic or financial states, but the plain GARCH model cannot deal with this.
To consider this state heterogeneity in the volatility process, researchers have developed state-heterogeneity GARCH-type models---for example, Markov-switching GARCH \citep{bauwens2010theory, bauwens2014marginal, gray1996modeling, haas2004new,  hamilton1994autoregressive, klaassen2002improving}, GJR-GARCH \citep{glosten1993relation}, and QR-GARCH \citep{nyberg2012risk} models.
Their empirical studies support the existence of state heterogeneity in financial volatility. % and the advantage of unifying time-and state-domain information to account for the volatility dynamics.

GARCH family models generally use daily return information to determine daily volatility levels, but daily return squares provide limited information about current volatility levels \citep{andersen1998answering}. 
Therefore, the data period should be long enough to enjoy the large-sample asymptotic properties of estimator. However, structural breaks in long-time-period data may deteriorate the estimation quality and the requirement for long-time-period data hinders investigation of short-term market dynamics. 
State heterogeneity models are severely limited in their exposure to the aforementioned issues because the number of parameters increases in proportion to the number of states, and the data are split among states.
Recently, widely available financial big data have shed light on this issue.
For example, thanks to advances of technology, high-frequency financial data are available, and we can accurately estimate volatility with relatively short-time-period data.
In particular, researchers have modeled high-frequency data based on continuous-time It\^o processes and proposed procedures for estimating realized volatility.
Examples include multi-scale realized volatility \citep{zhang2006efficient, zhang2011estimating}, pre-averaging realized volatility \citep{jacod2009microstructure}, quasi-maximum likelihood estimator (QMLE; \citealp{ait2010high, xiu2010quasi}), kernel realized volatility \citep{barndorff2008designing}, and robust pre-averaging realized volatility \citep{fan2018robust}. 
\cite{renault2011causality} suggested the  endogenous trading time robust realized volatility, and  
\cite{liu2018estimating} demonstrated that the pre-averaging estimator is robust for the zero-duration high-frequency data. 
The availability of these efficient realized volatility estimators had made a great impact on the volatility modeling and analysis.
For example, in regard to the modeling aspect, researchers have tried to bridge the gap between the discrete-time volatility model and continuous-time process \citep{kallsen1998option, nelson1990arch, wang2002asymptotic}. 
For the volatility dynamics analysis, realized volatility is employed as an innovation, which helps to improve estimation and prediction performance \citep{cerovecki2019functional, engle2006multiple, Kim2020Overnight, shephard2010realising, song2021volatility, tao2011large, visser2011garch}.
Recently, \cite{kim2016} introduced the unified continuous volatility process (unified GARCH-It\^o model) to provide a mathematical background for using high-frequency data in the GARCH model estimation. 
They showed that incorporating high-frequency financial data improves parameter estimation performance and helps analyze low-frequency market dynamics.  %\citep{visser2011garch}.
See also \citet{kim2016statistical, kim2019factor}.
In this manner, some state heterogeneous volatility models also incorporate high-frequency data.
For instance, researchers employed realized volatility in regime-switching ARMA-GARCH \citep{zhang2015improving}, two-stage three-state FIGARCH \citep{shi2015modeling}, realized GARCH \citep{hansen2012realized}, and multivariate Markov regime-switching GARCH \citep{lai2017multivariate} models.
These studies reported the usefulness of high-frequency data in analyzing state heterogeneity in low-frequency financial volatility. 
The success of previous studies have increased interest in developing volatility models that provide a mathematical background for using high-frequency data to analyze low-frequency volatility dynamics.

To examine and account for state heterogeneity in low-frequency volatility dynamics based on high-frequency financial data, we propose a novel volatility model based on a continuous-time It\^o diffusion process whose instantaneous volatility process evolves depending on the state variables.
%It integrates time- and state-domain information by embedding homogeneous continuous instantaneous volatility dynamics and heterogeneous integrated dynamics.
In particular, its instantaneous volatility process is continuous with respect to time and has a homogeneous process  during  each low-frequency period. 
In contrast, the process varies with the state  at  each low-frequency period to capture the low-frequency market dynamics. 
Consequently, its integrated volatility process has a form of the famous regime switching GARCH model.
% and to capture the low-frequency market dynamics, the instantaneous volatility has a homogeneous process during each low-frequency period.
% On the other hand, each low-frequency period has a stat dependent instantaneous volatility process.
The model is called the state heterogeneous GARCH-It\^o (SG-It\^o) model.
To estimate model parameters, we suggest a quasi-maximum likelihood estimation procedure based on the high-frequency data and establish its asymptotic theories.
%We also show that the established statistical inferences can be generally applied to the wide range of state heterogeneity volatility models.
Furthermore, to test state heterogeneity in low-frequency volatility, we introduce a Wald test-type hypothesis testing procedure.
The results of empirical studies suggest the existence of leverage, trading volume or investor attention, market illiquidity, stock market comovement, and post-holiday effect on S\&P 500 index volatility.
However, these state heterogeneities are not revealed with the same period of low-frequency data, because of its inefficiency.
More details are provided in Section \ref{sec empirical study}.

%The proposed SG-It\^o model bridges a gap between low- and high-frequency data based modeling by allowing the state heterogeneity in continuous volatility processes.
% One of the main contributions of this paper is to bridge a gap between low- and high-frequency data based modeling by allowing the state heterogeneity in volatility process under continuous It\^o diffusion process.
%This study contributes in the field of financial econometrics by providing the rigorous mathematical background to utilize the well-performing realized volatility estimator for analyzing low-frequency volatility dynamics such as state heterogeneity.

% This provides rigorous mathematical background to utilize the well-performing realized volatility estimators in low-frequency market dynamics.
 %This study contributes in the field of high-frequency finance by suggesting new framework enables to integrate high-frequency time- and low-frequency state-domain information for analyzing financial volatility and 

%We justify our framework by comparing one-day-ahead volatility prediction. 
%The prediction results show that continuous and state heterogeneity SG-It\^o model has superior predictive ability.
%The results imply that taking state-domain information into account for time-series model improves model specification and the SG-It\^o model effectively integrates information from time- and state-domains.

The rest of the paper is organized as follows. 
Section \ref{sec2} introduces the SG-It\^o model and illustrates properties of instantaneous and integrated volatility processes. 
Section \ref{sec3} presents the quasi-maximum likelihood method and establishes its asymptotic theories. 
Section \ref{sec4} suggests a hypothesis testing procedure. 
Section \ref{sec5} provides the results of numerical studies, including simulation and empirical studies. Section \ref{sec6} concludes the paper. 
The proofs are provided in the Appendix.

\section{State heterogeneous GARCH-It\^o (SG-It\^o) model}\label{sec2}

\subsection{State heterogeneity in discrete-time volatility processes}\label{sec 2.1}

State heterogeneity in financial volatility has long been discussed as the key feature of market dynamics \citep{lamoureux1990persistence}.
To account for state heterogeneity in the volatility process, researchers have proposed various form of regime-switching GARCH (RS-GARCH) models.
For example, \cite{hamilton1994autoregressive} applied the Markov-switching approach to build the state heterogeneous GARCH process.
See also \citet[2014]{bauwens2010theory}, \cite{gray1996modeling}, \cite{haas2004new}, and \cite{klaassen2002improving}.
\cite{glosten1993relation} introduced the GJR-GARCH model, which reflects the well-known leverage effect \citep{black1976studies, christie1982stochastic, figlewski2000leverage, tauchen1996volume}.
Taking the state of the business cycle into account exogenously, \cite{nyberg2012risk} introduced the regime-switching GARCH-M model (QR-GARCH) model to examine the state-dependent risk-return relationship.

A regime-switching model is characterized by the joint process of historical log price \{$X_n$\} and state variable \{$s_n$\}. 
Specifically, a process of the state variable $s_t$ and an evolution process of $X_t$ for a given state identify the model structure \citep{lange2009introduction}.
In the case of the RS-GARCH model, the conditional volatility process depends on the sigma field generated by \{$X_n$\} ($\mathcal{F}^{x,L}_n = \sigma(X_n, X_{n-1}, X_{n-2}, \ldots)$) and \{$s_n$\} ($\mathcal{F}^{s}_n = \sigma(s_n, s_{n-1}, s_{n-2}, \ldots)$), where $n \in \mathbb{N}$ and $\mathbb{N}$ is the set of all non-negative integers. 
A general discrete-time RS-GARCH model is described as follows:
\begin{align}\label{eq RS-GARCH}
\begin{split}
   &X_n - X_{n-1} = \mu + \sqrt{h_{n}(\theta^{s,L})}\epsilon^L_n,\\
   &h_{n}(\theta^{s,L}) = \omega^L_i + \gamma^L_i h_{n-1}(\theta^{s,L})  + \beta^L_i \zeta_{n-1}^2,
\end{split}
\end{align}
where $\theta^{s,L} = (\omega^L_i, \gamma^L_i, \beta^L_i)$ is a model parameter for the state indicator $i = 1,2,\ldots,D$, $D$ is the number of states, $\mu$ is a drift, $\zeta_n = X_n - X_{n-1} - \mu$, and random error $\epsilon^L_n$ satisfies $E\left[\epsilon^L_n|\mathcal{F}^L_{n-1}\right] = 0$ and $E\left[\left(\epsilon^L_n\right)^2|\mathcal{F}^L_{n-1}\right] = 1$ a.s. for $\mathcal{F}^L_{n-1} = \mathcal{F}^{x,L}_{n-1} \cup \mathcal{F}^{s}_n$.
For the RS-GARCH model,  the model parameters vary with the state, so the state variable $s_t$ plays a key role.
The state variable $s_t$ may have variety of forms, and the assumption for the state process distinguishes the model.
For example, the Markov-switching GARCH model has a latent Markov state process, whereas the GJR- and QR-GARCH models employ exogenous state variables.

\subsection{State heterogeneous GARCH-It\^o model}\label{sec 2.2}

In the study of discrete-time market dynamics, a long time period of data is needed to obtain consistent estimation results.
However, the long period of data is prone to exposure to the structural break issue, especially for RS-GARCH-type models, because of their complexity. 
Recently, realized volatility estimators based on high-frequency financial data have been well developed \citep{ait2010high, barndorff2008designing, jacod2009microstructure, xiu2010quasi, zhang2006efficient}.
\citet{kim2016} showed improvement of parameter estimation efficiency by using realized volatility estimators as the estimation proxy.
Therefore, we hypothesize that (1) the state heterogeneity exists in financial volatility process and (2) using high-frequency data facilitates its analysis.
For hypothesis testing, a model that enables to utilize realized volatility estimators in the analysis of low-frequency state heterogeneity is required.
This section introduces a novel continuous-time volatility model whose instantaneous volatility process varies with a discrete state process. % whose instantaneous volatility process varies with the discrete state process. % and its corresponding integrated volatility structure should be suggested.
 
%To utilize realized volatility estimators in the volatility model inference, continuous volatility process and corresponding integrated volatility structure should be suggested.
%Despite its importance, no research bridges the gap between the discrete RS-GARCH model and continuous-time volatility process.
%In this study, we newly introduces continuous-time volatility model whose instantaneous volatility process varies with the discrete state process.
% Two identifiers of new state heterogeneity model (i.e. state process and volatility process for given state) are made as follows.
For the state variable process $s_n$, we consider discrete-time exogenous variables that determine the state of volatility process. % in terms of its GARCH properties.
The term $exogenous$ comes from the independence assumption between the state variable $s_n$ and price process, which describe the unilateral influence of the state on the volatility process. 
%This assumption facilitates the model specification and helps to circumvent a path dependence problem.
% Note that any economic and financial status or even non-econmic events can be a state.
% For example, leverage effect can be modeled by assigning $s_n = 1$ after negative return day and $s_n = 0$ otherwise. 
% Also, the day-of-week effect can be modeled by assigning $s_n = 1$ for specific day-of-week and $s_n = 0$ otherwise.
For simplicity, this study deals with the binary state by assuming \{$s_n$\} as a binary process.
%The instantaneous volatility process for the given state is developed based on an approach of \cite{kim2016} as follows.
Let $\mathbb{R}^+ = [0,\infty]$ and $t\in \mathbb{R}^+$. 
We define a state heterogeneity volatility model with a continuous-time It\^o process as follows.

\begin{definition}\label{def model} 
For $t\in\left(n-1,n\right]$, we call a log stock price $X_t$ follows an SG-It\^o model if it satisfies
\begin{align}
\begin{split}\label{eq hfmodel}
    &dX_t=\mu dt+ \sigma_t dB_t, \quad 	 \sigma_t = (1-s_{n})\sigma_{1,t}   + s_{n} \sigma_{2,t} , \\
    &\sigma_{i,t}^2=\sna+(t-n+1)\{\omega_i+(\gamma_i-1)\sna\}+\beta_i \left(\int_{n-1}^{t} \sigma_{i,s} dB_{s}\right)^2\ \text{for}\ i\in\{1,2\},\\
    %&\sigma_n = (1-s_{n})\sigma_{1,n} + s_{n}\sigma_{2,n},
    % &s_n = 
    % \begin{cases}
    % 0, \mbox{with probability } p_n\\
    % 1, \mbox{with probability } 1-p_n
    % \end{cases},
\end{split}
\end{align} 
where $X_t$ is a log stock price, $\sigma_{1,t}^2$ and $\sigma_{2,t}^2$ are volatility processes adapted to $\mathcal{F}_t^x = \sigma(X_s: s \le t)$, $B_{t}$ is the standard Brownian motion  with respect to a filtration $\mathcal{F}_t^x$, and $\theta =  (\wa,\wb,\ga,\gb,\ba,\bb  )$ are model parameters.
\end{definition}

For the SG-It\^o model, instantaneous volatility has a continuous-time state heterogeneous process defined at all times $t$. %reflecting discrete state process defined at integer time point $n$.
In particular, during the current low-frequency period $t\in\left(n-1,n\right]$, the instantaneous volatility $\sigma_{s_n+1,t}^2$ has a homogeneous continuous-time It\^o process depending on the current state variable $s_n$.
During the next low-frequency period $t\in\left(n,n+1\right]$, the instantaneous volatility $\sigma_{s_{n+1}+1,t}^2$ evolves from the end of the previous instantaneous process $\sigma_{s_n+1,n}^2$ while their evolving process is determined by the next period state variable $s_{n+1}$.
This model deals with the low-frequency state heterogeneity in volatility by state-varying coefficients.
For example, for $s_n = 0$, we have $(\omega_i, \gamma_i, \beta_i) = (\wa,\ga,\ba)$ and $\sigma_{t}^2 =  \sigma_{1,t}^2$, whereas for $s_n = 1$, we have $(\omega_i, \gamma_i, \beta_i) = (\wb,\gb,\bb)$ and $\sigma_{t}^2 =  \sigma_{2,t}^2$.
Moreover, the current level of volatility depends on past volatility due to its recursive structure.
Thus, the model is uniquely identified by the path of the state variables because the sequential order of the states differentiates the volatility process.
Accordingly, it can handle the regime shift in volatility with the corresponding state variable.
Since the model can incorporate any state process, it allows us to test a given exogenous state.
This is a distinguishing feature of the SG-It\^o model compared to the single-regime model.
When the states are homogeneous ($s_n = s_{n-1} = \cdots = s_1$), the model returns to a single-regime model, the unified GARCH-It\^o model \citep{kim2016}.
% Note that when the states are homogeneous ($s_n = s_{n-1} = \cdots = s_1$), the SG-It\^o model returns to the unified GARCH-It\^o model \citep{kim2016}.
That is, the unified GARCH-It\^o model is a special example of the SG-It\^o model.
More details are provided in Appendix \ref{appendix garchito}.

% For example, if $s_n = 0$, $\sigma_t = \sigma_{1,t}$ for $t\in\left(n-1,n\right)$ and $\sigma_n = \sigma_{1,n}$
% for integer time point $t=n$ as well.
In this paper, we consider the case where $n\in\mathbb{N}$ denotes a day. 
In this case, $s_n$ is a daily state variable, and $\sigma_t^2$ for $t\in\left[n-1,n\right)$ denotes the intraday volatility on day $n$. 
Definition \ref{def model} now implies that intraday volatility on day $n$ evolves from close volatility on day $n-1$ reflecting the contemporaneous log stock price, while the evolving process depends on the daily state variable $s_n$.

\subsection{Integrated volatility for the SG-It\^o model}
%In the analysis of continuous volatility process, not only instantaneous volatility but also integrated volatility $\nint \si^2_t dt$ is often examined.
%The integrated volatility structure is important in this study since this study bridges the gap between continuous-time process and discrete-time volatility model.
%We achieve it by showing expected integrated volatility has a RS-GARCH-like structure.

This study aims to investigate the low-frequency market dynamics, so the integrated volatility structure over the low-frequency period is important. 
Moreover, the integrated volatility process will be used in the parameter estimation procedure.
In this section, we study properties of the integrated volatilities.

\begin{theorem}\label{thm1}(a) Under the SG-It\^o framework, integrated volatility on state $i\in\{1,2\}$ can be decomposed into \{$\mathcal{F}_{n-1}^{x,L},\mathcal{F}_{n-1}^{s}\}$-adapted process and martingale difference as follows: 
\begin{align*}
    &\nint \si_{i,t}^2dt=h_{i,n}(\theta)+\xi_{i,n} \text{ a.s.},
\end{align*}
where
\begin{align*}
    &h_{i,n}(\theta) = H_{c,i}(\theta)+H_{\beta,i}(\theta)\sna,\  \xi_{i,n}=2\nint(e^{(n-t)\beta_i}-1)\int_{n-1}^{t}\si_{i,s}dB_{s}\si_{i,t}dB_{t},\\
    &\Hic=\betai^{-2}(e^{\betai}-1-\betai)\wi,\  \Hib=(\gi-1)\betai^{-2}(e^{\betai}-1-\betai)+\betai^{-1}(e^{\betai}-1).
\end{align*}
(b) Let $\mathcal{F}_{n-1} = \mathcal{F}_{n-1}^{x}\cup\mathcal{F}_{n}^s$. Then, for given $s_n$ and $s_{n-1}$, the conditional expected integrated volatility $E\left[\nint \si_{t}^2dt\sfilta\right]=h_n(\theta)$ a.s. is represented by
\begin{align}
\begin{split}\label{eq integrated SG-Ito}
    \hn&=\yaya(\wah+ \gah \hna+\bah \Zna)+\yayb(\wbh+\gbh \hna+ \bbh\Zna)\\
    &\quad+\ybya(\wch+ \gch  \hna+\bch\Zna)+\ybyb(\wdh+ \gdh\hna+ \bdh\Zna),
\end{split}
\end{align}
where
\begin{align*}
    &\yaya=\yab\yaa, \yayb=\yab\yba,\ybya=\ybb\yaa,\ybyb=\ybb\yba,\\
    &\theta^h = \{\wah,\wbh,\wch,\wdh,\gah,\gbh,\gch,\gdh,\bah,\bbh,\bch,\bdh\},\\
    &\omega_{ii}^h=(1-\gamma_i)\Hic+\omega_i\Hib,\ \gamma_{ii}^h=\gamma_i,\ \beta^h_{ii}=\beta_{i}\Hib,\\
    &\omega_{ij}^h=\Hjc-\gamma_i\Hic(\Hjb/\Hib)+\omega_i\Hjb,\ \gamma_{ij}^h=\gamma_i(\Hjb/\Hib),\ \beta_{ij}^h=\beta_i\Hjb,\\
    &Z_n = (1-s_{n})Z_{1,n}+s_{n}Z_{2,n},\ Z_{i,t} = \int_{t-1}^{t} \sigma_{i,s} dB_{s}\ \, \text{for}\ i\in\{1,2\}.
\end{align*}
\end{theorem}

Theorem \ref{thm1}(a) shows that integrated volatility on state $i$ is decomposed into GARCH volatility $h_{i,n}(\theta)$ and martingale difference $\xi_{i,n}$. 
This decomposition plays a prominent role in the subsequent theorems, and we show that the theorems can be established for any process that satisfies the decomposition in Theorem \ref{thm1}(a).
Theorem \ref{thm1}(b) indicates that the expected integrated volatility $h_{n}(\theta)$ follows a four-state RS-GARCH(1,1) structure.
In particular, the integrated form of the model parameter $\theta^h$ is determined by the product of $s_n$ and $s_{n-1}$, so the integrated volatility depends on both current and previous states.
In the sense that its integrated volatility has a RS-GARCH-like structure, the SG-It\^o model has an instantaneous volatility process that characterizes the RS-GARCH models. % and fills the gap between the continuous-time modeling and the discrete-time regime-switching volatility structure.
We note that this general model allows to incorporate and extend existing regime-switching volatility frameworks by employing a suitable state process.
For example, this model illustrates the Markov-switching GARCH model with latent Markov state process and the GJR- and QR-GARCH models with observed exogenous state processes. 

%\b{This study mainly deals with observed state processes for the convenience.}
 In this paper, we mainly deal with observed state processes.
In practice, however, future state is often unobservable. 
% when the forecasting stretches beyond one-day-ahead.
For instance, day-of-week or previous day return state is available at the beginning of the day, whereas daily trading volume or market illiquidity is not available until the end of the day.
We note that the SG-It\^o model does not require the observability of the state variable $s_{n}$.
 For unrevealed $s_{n}$, Proposition \ref{prop1} suggests that we can estimate the expected integrated volatility with state transition probability. 
% For unrevealed $s_{n}$, Proposition \ref{prop1} suggests that we can make a one-step-ahead integrated volatility forecast with state transition probability. 

\begin{proposition}\label{prop1} For unrevealed $s_n$ and given $s_{n-1}$, we have
\begin{align*}
    E\left[\nint \si_{t}^2dt\Big|\mathcal{F}_{n-1}^x, \mathcal{F}_{n-1}^s\right]&=p_{11,n}\yab(\wah+\gah\hna+\bah\Zna)\\
    &\quad+p_{12,n}\yab(\wbh+\gbh\hna+\bbh\Zna)\\
    &\quad+p_{21,n}\ybb(\wch+\gch\hna+\bch\Zna)\\
    &\quad+p_{22,n}\ybb(\wdh+\gdh\hna+\bdh\Zna) \text{ a.s.},
\end{align*}
where $p_{ij,n} = p(s_{n}=j-1|s_{n-1}=i-1)$ for $i,j \in \{1,2\}$.
\end{proposition}

In practice, $Z_n$'s are not observable due to the drift term $\mu$, thus to predict the future volatility, we first need to estimate $\mu$ using the sample mean of the daily log-returns. 
The martingale convergence theorem shows that the sample mean of daily log-return converges to $\mu$.

\section{Estimation procedure} \label{sec3}
 
 \subsection{Model setup}
In this paper, we assume that the true log price process follows the SG-It\^o model as described in Definition \ref{def model}.
We also distinguish the low- and high-frequency data as follows.
The low-frequency data signify the log price observed at integer time points $t = 0,1,2,\ldots$ and we assume that the true low-frequency log prices, $X_0, X_1,\ldots$, are observed.
At the same time, high-frequency data indicate the log price observed at time points between integer time points, which are denoted by $t_{n,m}$ for $n = 0,1,\ldots,N$ and $m =  1,\ldots,M_n-1$ for each $n$ and satisfy $n-1 = t_{n,0} < t_{n,1} < \cdots < t_{n,M_n} = t_{n+1,0} = n$.
In the high-frequency finance, the observed price $Y_{t_{n,m}}$ is contaminated by micro-structure noises.
To reflect this, we assume that $Y_{t_{n,m}}$ is composed of the true price $X_{t_{n,m}}$ and micro-structure noise $\epsilon_{t_{n,m}}$ as follows:
$$
    Y_{t_{n,m}} = X_{t_{n,m}}+\epsilon_{t_{n,m}},
$$
where $\epsilon_{t_{n,m}}$ is independent with the price and volatility process and i.i.d. with mean zero and standard deviation $\sigma_\epsilon$.
For the low-frequency data, we can estimate drift easily by calculating the sample mean of the low-frequency data.
For the high-frequency based realized volatility estimator, the effect of drift is asymptotically negligible.
Therefore, for simplicity, we assume $\mu = 0$ in Equations \eqref{eq RS-GARCH} and \eqref{eq hfmodel}.
This is not a necessary condition but allows us to focus more on developing and analyzing volatility processes.
% This allows us to focus more on developing and analyzing volatility processes.

\subsection{Quasi-maximum likelihood estimation}

For convenience, we first review some notations and definitions.
Unless stated otherwise, limits are taken as $N,M\rightarrow\infty$, where $M=\sumn M_n/N$. 
Let $\xrightarrow{\ p\ }$ and $\xrightarrow{\ d\ }$ be convergence in probability and distribution, respectively. 
The $L_p$ norm of a random variable Z is denoted by $\lpnorm{Z}=\left(E\left[\vert Z \vert ^p \right]\right)^\frac{1}{p}$.
Finally, $\maxnorm{X} = \max_{j,k}\left\vert X_{j,k}\right\vert$ for a matrix $X = (X_{j,k})_{j,k = 1,\ldots, q}$ and $\maxnorm{x} = \max_{j}\left\vert  x_{j}\right\vert$ for a vector $x = (x_1,\ldots,x_q)$.
$C$'s present a positive generic constant whose values can be changed from appearance to appearance, free from $\theta$, $N$, and $M_n$.

For statistical inferences, we apply a quasi-maximum likelihood estimation procedure to the integrated volatility process.
Theorem \ref{thm1}(a) suggests that integrated volatility over the $n$th period is decomposed into the GARCH volatility $\hn$ and martingale difference. 
The well-developed martingale convergence theorem indicates that the integrated volatility converges to $\hon$ as $N\rightarrow\infty$, so the integrated volatility can be a good proxy of $\hon$.
Unfortunately, integrated volatility is not observed, so we need to estimate it using the observed noisy high-frequency data. 
There are well-performing realized volatility estimators such as multi-scale realized volatility \citep{zhang2006efficient, zhang2011estimating}, pre-averaging realized volatility \citep{jacod2009microstructure}, and kernel realized volatility \citep{barndorff2008designing}, which have the optimal convergence rate $M^{-1/4}_n$.
We adopt the pre-averaging realized volatility in the numerical studies. 
%In the proof of Theorem \ref{thm3}, we prove that using realized volatility estimator instead of daily return square decrease asymptotic variance of the parameter estimates.

% \clearpage

% \begin{align*}
% \Lhn - \Ln = -\frac{1}{2N}\sumn \frac{\xi_n}{\hn}.
% \end{align*}
% Let
% \begin{align*}
% \frac{1}{4N}\sumn \left(\frac{\xi_n}{\hn}\right)^2 \xrightarrow{\ p\ }V^1.
% \end{align*}
% Then, by martingale CLT,
% \begin{align*}
% \Lhn - \Ln \xrightarrow{\ d\ } N\left(0,V^1\right).
% \end{align*}
% Let
% \begin{align*}
% \widetilde{V}_1(\theta) &= \frac{1}{4N}\sumn \left(\frac{\hon + \xi_n - \hn}{\hn}\right)^2 \\
% &=\frac{1}{4N}\sumn \left(\frac{\hon + \xi_n}{\hn} -1 \right)^2.
% \end{align*}
% We can show that $\widetilde{V}_1(\theta)$ is minimized at $\theta = \to$.
% We conduct a variance ratio test to test $\theta = \to$ using
% \begin{align*}
% F = \frac{\hat{V}_1(\to)}{\hat{V}_1(\theta)} \quad\text{where}\quad
% \hat{V}_1(\theta) = \frac{1}{4N}\sumn \left(\frac{RV - \hn}{\hn}\right)^2.
% \end{align*}

% \clearpage

Let $\theta = (\wa, \wb, \ba, \bb, \ga, \gb) \in \Theta$ with the true value $\to = (\woa, \wob, \boa, \bob, \goa, \gob) \in \Theta$ for the compact parameter space $\Theta$. 
Then, for the given state variable $s_{n}$, the quasi-maximum likelihood function is defined as follows:
\begin{align}\label{eq qml function}
    &\Lhnm=-\frac{1}{2N}\sumn\left[ \log(\hn)+\frac{RV_n}{\hn}\right],
\end{align}
where $RV_n$ is the realized volatility estimator constructed based on high-frequency data during the $n$th period and $\hn = \yaa h_{1,n}(\theta) + \yba h_{2,n}(\theta)$ is state-heterogeneous GARCH volatility presented in Theorem \ref{thm1}(b).
% For simplicity, we drop the terms related to state probability $p$ by employing observed exogenous state process.
The estimation procedure can be easily generalized to the other forms of state processes with a suitable quasi-likelihood function.
The QMLE $\th$ is 
 $$
 \th = \mathop{\mathrm{argmax}}_{\tinT} \Lhnm.
 $$
%\b{We note that coefficients of each state cannot be estimated separately due to the sequential structure of $\hn$.}
To investigate the asymptotic behaviors of QMLE $\th$, we need the following technical conditions. 

\begin{assumption}\label{ass1} ~
\begin{itemize}\itemsep -0.05in
\item[(a)] Let the parameter space $\Theta=\{\theta = (\wa,\wb,\ga,\gb,\ba,\bb):\omega_l<\omega_i<\omega_u,\  \gamma_l<\gamma_i<\gamma_u,\ \beta_l<\beta_i<\beta_u,\ \wlh<\omega_{ij}^h<\wuh,\ \glh<\gamma_{ij}^h<\guh,\ \blh<\beta_{ij}^h<\buh,\ \gamma_{ij}^h+\beta_{ij}^h<1\}$ for $i,j\in\{1,2\}$, where $\omega_l,\omega_u,\gamma_l,\gamma_u,\beta_l, \beta_u,\wlh, \wuh,\glh, \guh,\ \blh, \buh$ are known positive constants.
\item[(b)] $\supn E(|\xi_{i,n}|^{1+\delta}) < \infty$  for $i\in\{1,2\}$ and some $\delta>0$.
% \item[(c)]For any $\ninN$, $\ltwonorm{\frac{1}{\hn}\dhj}\le C$, $\ltwonorm{\frac{1}{\hn}\ddhjk}\le C$, and
% $\ltwonorm{\frac{1}{\hn}\dddhjkl}\le C$ for any $j,k,l\in\{1,2,3,4,5,6\}$, where $\theta=(\theta_1,\theta_2,\theta_3,\theta_4,\theta_5,\theta_6)=(\wa,\wb,\ga,\gb,\ba,\bb)$.
\item[(c)]$\frac{E[Z_n^4\sfilta]}{\hon^2}\le C$ a.s. for any $\ninN$.
 \item[(d)]$\{\xi_{i,n},Z_{i,n}, s_n\}$ is a stationary and ergodic process. 
% \item[(d)]$\{\xi_n,\Zn, s_n\}$ is a stationary and ergodic process, where $\xi_n = \yaa \xi_{1,n} + \yba \xi_{2,n}$.
\end{itemize}
\end{assumption}

\begin{remark}
Assumption \ref{ass1} is required to handle the low-frequency part.
Assumption \ref{ass1}(b) is the sufficient condition for the uniform integrability of martingale difference process.
The uniform integrability is a necessary condition to show the boundedness of derivatives of the quasi-likelihood functions, which is required to obtain the consistency of $\th$. 
% Assumption \ref{ass1}(c) is the bounded moment condition for the first, second, and third derivatives of the GARCH volatility $\hn$. 
Assumption \ref{ass1}(c) is the finite fourth moment condition. 
Because the target parameter is the second moment, the finite fourth moment condition is not strong at all to obtain the convergence rate $N^{-1/2}$ (see also \cite{lee1994asymptotic}).
 Assumption \ref{ass1}(d) is only required to derive asymptotic normality of $\th$.
However, this condition is not an obvious result under the SG-It\^o model.
It is an interesting theoretical problem to investigate conditions which imply   Assumption \ref{ass1}(d) under the SG-It\^o model. 
We leave this for the future study.
\end{remark}

\begin{remark}
The state process should be stationary and ergodic.
It includes state processes that are generally applied in existing switching models.
For example, any multinomial variables of ergodic probabilities are included.
Specifically, the state variable $s_t$ satisfies
$$
s_t|\Omega_{t-1} \sim Bernoulli(p_t),\quad
p_t = E_{t-1}(s_t = 1) = \Phi(\pi_t),
$$
where $\Phi(\cdot)$ is a link function and $\pi_t$ is explanatory variables.
Then, under some stationary condition for $\pi_t$, the state variable $s_t$ is  a stationary and ergodic process.
\end{remark}

\begin{assumption}\label{ass2}~
\begin{itemize}\itemsep -0.05in
\item[(a)]Assume $C_1M\le M_n \le C_2M$, $\sup_n \sup_{1\le m\le
M_n}|t_{n,m}-t_{n,m-1}|=O(M^{-1})$, and $N^2M^{-1}\rightarrow 0$ as $N,M\rightarrow\infty$.
\item[(b)]$\supn\RVsig\le CM^{-\frac{1}{4}}$.
\item[(c)]For any $\ninN$, $E[RV_n\sfilta]\le CE\left[\isign\sfilta\right]+C$ a.s.
\end{itemize}
\end{assumption}

\begin{remark}
Assumption \ref{ass2} stands for the high-frequency part. 
Assumption \ref{ass2}(a) is a typical condition for realized volatility estimators. 
Assumption \ref{ass2}(b)--(c) can be obtained easily under some fourth moment conditions as discussed in \cite{tao2013fast},   \cite{kim2016asymptotic}, and \cite{kim2018large}.
 \end{remark}

% We now able to deduce following asymptotic results under the assumptions. 
Theorems \ref{thm2} and \ref{thm3} establish the consistency of $\th$ and its convergence rate, respectively.
\begin{theorem} \label{thm2} 
Under Assumption \ref{ass1}(a)--(b) and Assumption \ref{ass2}(a)--(b), we have $\th\rightarrow\to$ in probability.
\end{theorem}

\begin{theorem}\label{thm3}
 Under Assumption \ref{ass1}(a)--(c) and Assumption \ref{ass2}, we have $\maxnorm{\th-\to}=O_p(N^{-\frac{1}{2}}+M^{-\frac{1}{4}})$.
\end{theorem}

\begin{remark}
Theorem \ref{thm3} shows that the convergence rate of $\th$ has both high- and low- frequency-oriented components.
The rate $N^{-\frac{1}{2}}$ is due to the low-frequency part, which is the usual parametric convergence rate. 
The rate $M^{-\frac{1}{4}}$ is from the high-frequency volatility estimation related to Assumption \ref{ass2}(b), known as the optimal convergence rate of the realized volatility estimator with the presence of the micro-structure noise.
\end{remark}

Theorem \ref{thm4} derives asymptotic normality of $\th$ using stationary and ergodic assumptions.

\begin{theorem}\label{thm4} 
Suppose that Assumptions \ref{ass1} and \ref{ass2} are met and 
\begin{align*}
    &\frac{1}{4N}\sumn\left[\dht\dhtt\Big|_{\theta = \to} \hon^{-4} \xi_n^2\right]  \xrightarrow{\ p\ }V,\\
    &\frac{1}{2N}\sumn\left[\dht\dhtt\Big|_{\theta = \to} \hon^{-2} \right] \xrightarrow{\ p\ }W.
\end{align*} 
Then we have  
\begin{align*}
    \sqrt{N}(\th-\to)\xrightarrow{\ d\ } N\left(0,W^{-1}VW^{-1}\right).
    \end{align*}
\end{theorem}

Theorem \ref{thm4} demonstrates that the limiting distribution of QMLE is Gaussian with the variance $W^{-1}VW^{-1}$, where the matrices $V$ and $W$ are information and Hessian matrices, respectively. 
 Theorem \ref{thm4} implies that the quality of the integrated volatility estimator affects the variance of the parameter estimates.
To check the effect of employing integrated volatility estimators as the proxy, we consider the parameter estimation procedure using low-frequency data only.
For example, the QMLE $\th^L$ is obtained as follows: 
\begin{align}\label{eq lf qml function}
    \hat{Q}^{L}_{N}(\theta)=-\frac{1}{2N}\sumn \left[\log(\hn)+\frac{\zeta_n^2}{\hn}\right] \quad \text{and} \quad\th^L = \mathop{\mathrm{argmax}}_{\tinT} \hat{Q}^{L}_{N}(\theta),
\end{align}
where $\zeta_n = X_n - X_{n-1} - \mu$ is defined in Equation \eqref{eq RS-GARCH}.
Then, similar to the proof of Theorem \ref{thm4}, the asymptotic distribution of $\th^L$ can be derived as follows:
\begin{align*}
    \sqrt{N}(\th^L-\to)\xrightarrow{\ d\ } N\left(0,W^{-1}V^L W^{-1}\right),
\end{align*}
where
\begin{align*}
    \frac{1}{4N}\sumn\left[\dht\dhtt\Big|_{\theta = \to} \hon^{-4} \left(Z_n^2-\hon\right)^2\right]  \xrightarrow{\ p\ }V^L.
\end{align*}
From the above results, we can find that the estimation errors of the GARCH volatility in realized volatility (i.e., $\nint \sigma^2_t dt - \hon$) or daily return square (i.e., $Z^2 - \hon$) play a key role in the variance of parameter estimates.
We can easily find that  the estimation errors in realized volatility is smaller than that in  the daily return square. 
That is, compared to the daily return square, the sufficient information in realized volatility estimators reduces the parameter estimation error and produces accurate parameter estimates with a relatively short time period of data (see also \citet{kim2016}). 

To make inferences based on the asymptotic distribution derived in Theorem \ref{thm4}, we construct consistent estimators for $V$ and $W$ as follows:
 \begin{eqnarray*}
	&&\hat{V}=\frac{1}{4N}\sumn \left[\dht\dhtt\Big|_{\theta = \th}\ h_n(\th)^{-4} (RV_n -\ h_n(\th))^2\right],  \cr
	&&\hat{W}=\frac{1}{2N}\sumn \left[\dht\dhtt\Big|_{\theta = \th}\ h_n(\th)^{-2} \right].
\end{eqnarray*}
% where $h_n(\theta)$ is defined in Equation \eqref{eq integrated SG-Ito}.

To show the consistency of estimators, we make following additional assumptions.

\begin{assumption}\label{ass3}~
\begin{itemize}\itemsep -0.05in
\item[(a)] $\frac{E[Z_n^8\sfilta]}{\hon^4}\le C$ a.s. for any $\ninN$.
\item[(b)]$\supn\lfournorm{RV_n - \isign}\le CM^{-\frac{1}{4}}$.
\end{itemize}
\end{assumption}

\begin{remark}
Assumption \ref{ass3} is the finite eighth moment condition.
The parameters, $V$ and $W$, of interest are functions of fourth moments.  
Thus, to establish their asymptotic theorems, we need the finite eighth moment condition.
\end{remark}

\begin{proposition}\label{prop2}
 Under Assumptions \ref{ass1}--\ref{ass3}, we have $\hat{V}\rightarrow V$ and $\hat{W}\rightarrow W$ in probability.
\end{proposition}

%\begin{remark}
%$V$ and $W$ are the components of asymptotic variance of QMLE and, accordingly, their quantities are required to make statistical inferences.
%In this paper, we construct some hypothesis test in Section \ref{sec4} to test the heterogeneity of volatility processes.  
%We employ the consistent estimators $\hat{V}$ and $\hat{W}$. 
%Details can be found in Theorem \ref{thm5}.
%\end{remark}

\begin{remark}
Proposition \ref{prop2} shows the consistency of $\hat{V}$ and $\hat{W}$. 
We can obtain their convergence rates by imposing some additional condition.
For example, by assuming that $\{\hon, \xi_n ^2\}$ is a strong mixing sequence, Theorem 1.2 \citep{merlevede2000functional} shows that 
$\hat{V}$ and $\hat{W}$ have the convergence rate $N^{-1/2} + M^{-1/4}$. 
\end{remark}

\section{Testing state heterogeneity} \label{sec4}

The main purpose of this paper is to investigate state heterogeneity in the volatility process.
 In the previous section, we propose a state-heterogeneous diffusion process that can incorporate the low-frequency state process. 
Under the SG-It\^o model, state heterogeneity in the volatility process is illustrated by the state-varying parameters $\omega_i$, $\gamma_i$, $\beta_i$.
Therefore, we can test the state heterogeneity by conducting a hypothesis test under the null hypothesis statement  
$$H_0:  \wa = \wb,\ \ga = \gb,\ \ba = \bb .
$$
In this section, we construct a Wald test-type hypothesis testing procedure with the null hypothesis $H_0$ for the QMLE.
The rejection of the null hypothesis signifies that the external state distinguishes the model specification, which implies the existence of state heterogeneity in the volatility process.
% Generally, likelihood ratio test can be applied, but existing studies prove that quasi-likelihood ratio statistics may not follows asymptotic $\chi^2$ distribution under the null hypothesis. In particular, likelihood ratio statistic is distinguished from Wald statistic when the information matrix equality is not established (i.e. $V + W \ne 0$) as in our case. Accordingly, we apply Wald-type test in hypothesis testing.
Let $R$ be the $v\times u$ restriction matrix with full row rank.
% rank and $\hat{V}$ and $\hat{W}$ be the consistent estimator of $V$ and $W$, respectively. 
Theorem 5 defines the Wald-type statistic and establishes its limiting distribution.

\begin{theorem}\label{thm5} Under Assumptions \ref{ass1}--\ref{ass3} and the null hypothesis of $R\to = r$, we have 
\begin{align*}
    T_{N,M}= N(R\th-r)^T(R\hat{W}\hat{V}^{-1}\hat{W}R^T)^{-1}(R\th-r)\xrightarrow{\ d\ }\chi^2(v),
\end{align*}
where $\chi^2(v)$ indicates a chi-squared random variable with $v$ degrees of freedom. 
% where
% \begin{align*}
%     &\hat{V}=\frac{1}{4N}\sumn\left[\dht\dhtt\Big|_{\theta = \to} \hon^{-4} \xi_n^2\right],    &\hat{W}=\frac{1}{2N}\sumn\left[\dht\dhtt\Big|_{\theta = \to} \hon^{-2} \right].
% \end{align*}
\end{theorem}

 Theorem \ref{thm5} suggests that the asymptotic normality of $\th$ induces a Wald-type statistic that follows asymptotically $\chi^2$ distribution under the null hypothesis. 
% The consistent estimator of information matrix $V$ and Hessian matrix $W$ are suggested in the proof of Theorem \ref{thm5}. 
We can test the null hypothesis $H_0$ by setting 
\begin{align*}
R=\begin{pmatrix}
    1 & 0 & 0 & -1 & 0 & 0 \\
    0 & 1 & 0 & 0 & -1 & 0 \\
    0 & 0 & 1 & 0 & 0 & -1 \\
\end{pmatrix}\quad\text{and}\quad
r = (0,0,0)^T.
\end{align*}
Then the Wald-type statistic $T_{N,M}$ follows $\chi^2(3)$.
In the empirical study, we reveal the state heterogeneity in S\&P 500 index volatility by conducting the proposed Wald-type test. 
Details can be found in Section \ref{sec5}.

%In the simulation study, we estimate $T_{N,M}$ under the null and alternative hypothesis, respectively, and verify that its finite sample distribution gradually closes to the suggested limiting distribution and the power of the test increases as $N,M\rightarrow\infty$. 

\begin{remark}
Theorems \ref{thm2}--\ref{thm5} are established based on the decomposition of expected integrated volatility in Theorem \ref{thm1}(a),
$$
\nint \si_{i,t}^2dt=h_{i,n}(\theta)+\xi_{i,n} \text{ a.s.},
$$
and their results can be established for any instantaneous volatility process that satisfies Theorem \ref{thm1}(a).
The SG-It\^o model is one of the examples.
%Consequently, we provide rigorous mathematical background for using the well-performing realized volatility estimator in parameter estimation of a wide range of the RS-GARCH-family.
\end{remark}

\section{Numerical studies} \label{sec5}
\subsection{Simulation studies} \label{sec simulations}

To evaluate the relevance of asymptotic theories, we conducted simulation studies. 
We first simulated the log price process and assessed the finite sample performance of the suggested estimator $\th$.
The log stock price $X_{t_{n,m}}$ for $t_{n,m} = n-1+m/M$ was generated from the SG-It\^o model in Definition \ref{def model} with the following form:
\begin{align}\label{eq generate}
\begin{split}
    &dX_t=\mu dt+ \sigma_t dB_t, \quad 	 \sigma_t = (1-s_{n})\sigma_{1,t}   + s_{n} \sigma_{2,t} , \\
    &\sigma_{i,t}^2=\sigma_{[t]}^2+(t-[t])\{\omega_{0,i}+(\gamma_{0,i}-1)\sigma_{[t]}^2\}+\beta_{0,i} (X_t - X_{[t]}- (t-[t])\mu )^2,\\
    &  s_n = \mathbbm{1}\{(X_{n-1} - X_{n-2})<0\},
   % &Z_{i,t} = \sum_{s=t-1}^{t} \sigma_{i,s} dB_{s}\ \, \text{for}\ i\in\{1,2\},\,
\end{split}
\end{align}
where $\mathbbm{1}\{\cdot\}$ is an indicator function and $\to$ = $\tolist$ is the true model parameter.
To capture the leverage effect, we set $s_n=1$ if the previous day return is negative and $s_n=0$ otherwise.
The price process was generated under the null and alternative hypothesis, respectively.
Under the null hypothesis, the true parameter set is given by $\to = (0.15,0.15,0.2,0.2,0.1,0.1)$, whereas under the alternative hypothesis, the true parameter set is given by $\to = (0.15,0.165,0.2,0.22,0.1,0.11)$.
We set initial price $X_0 = 10$, initial instantaneous volatility $\sigma_{i,0}^2 = \frac{\omega^h_i(1-\beta^h_i-\gamma_i)+\beta_i \omega^h_i}{(1-\beta^h_i-\gamma_i)(1-\gamma_i)}$, and $\mu = 0$. 
We chose $N = 1000$ and $M = 23400$, corresponding to the stock price observed every second during four years.
The Euler scheme was applied to discritize continuous-time processes.
The observed price $Y_{t_{n,m}}$ was calculated as the sum of the true log price $X_{t_{n,m}}$ and the micro-structure noise $\epsilon_{t_{n,m}}$, where $X_{t_{n,m}}$ was generated from Equation \eqref{eq generate} and $\epsilon_{t_{n,m}}$ was generated from i.i.d. normal distribution with mean zero and standard deviation $\sigma_\epsilon = 0.01$. 
For realized volatility estimator, we employed the pre-averaging method \citep{christensen2010pre, jacod2009microstructure}, presented as follows:
\begin{align*}
    &RV_n=\frac{1}{\phi_{K}(f)}\frac{M}{M-K}\sum_{k=1}^{M-K+1}(\overline{Y}(t_k)^2-\frac{1}{2}\hat{Y}_k),\\
    &\overline{Y}(t_k) = \sum_{i=1}^{K-1}f\left(\frac{i}{K}\right)[Y_{t,k+i}-Y_{t,k+i-1}],\ \phi_{K}(f) = \sum_{i=1}^{K}f\left(\frac{i}{K}\right)^2,\\
    &\hat{Y}_k = \sum_{i=1}^{K}\left(f\left(\frac{i}{K}\right)-f\left(\frac{i-1}{K}\right)\right)^2(Y_{t,k+i}-Y_{t,k+i-1})^2,
\end{align*}
where $K=\left[\sqrt{M}\right]$ is the tuning parameter that determines the number of observations used for the pre-averaging step and $f(x) = \min(x,1-x)$. 
Using the generated stock prices, we estimated the realized volatility and calculated $\th$ using the QMLE method in Section \ref{sec3}. 
The simulation procedure was repeated 1000 times.

\begin{figure}[!ht]
\centering
\includegraphics[width=15cm, height = 10cm]{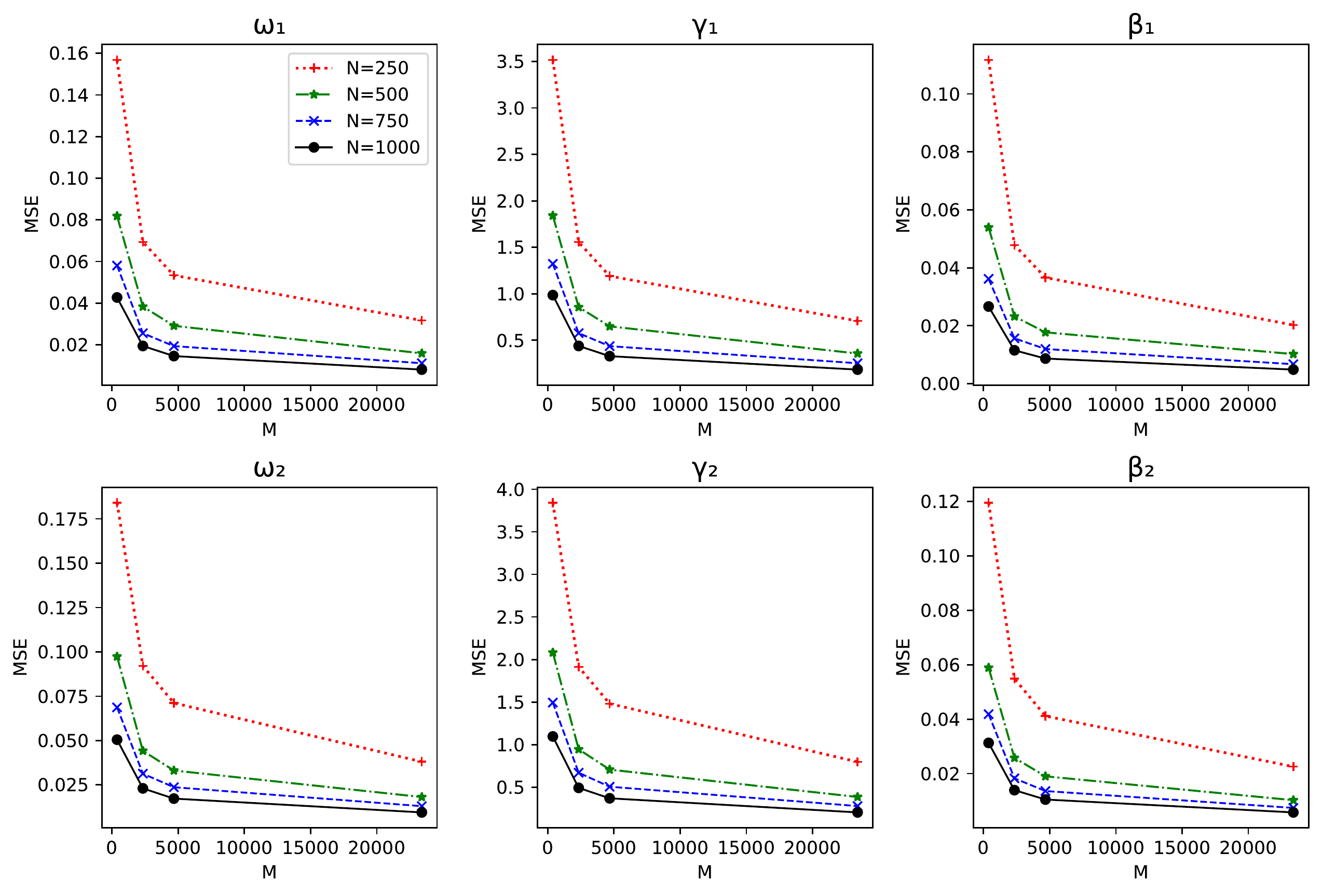}
\caption{MSEs of parameter estimates of the SG-It\^o model}
\begin{flushleft}
\small{\textit{Notes.} This figure illustrates MSEs of parameter estimates of the SG-It\^o model based on data simulated from the SG-It\^o model with $N = 250, 500, 750, 1000$ and $M = 390, 2340, 4680, 23400$.}
\end{flushleft}
\label{Figure-2}
\end{figure}

We first examined the effect of period and frequency of the data on parameter estimation.
The accuracy of model parameter estimation is expected to be improved by longer period and higher frequency data. 
To verify this, we generated additional data sets by resampling the entire data. 
Specifically, we collected first 250, 500, 750, and 1000-day data for $N = 250, 500, 750$, and $1000$, respectively. 
We also collected one of every 60, 10, and 5 data in each day corresponding to 1-minute ($M = 390$), 10-second ($M = 2340$), and 5-second data ($M = 4680$), respectively. 
Figure \ref{Figure-2} provides mean squared errors (MSEs) of the estimator $\th$ for varying $N$ and $M$. 
From Figure \ref{Figure-2}, we find that the use of longer period and higher frequency data significantly improves estimation performance, which supports the theoretical findings in Section \ref{sec3}.

\begin{center}
    \textbf{[Figure \ref{Figure-2} inserted about here]}
\end{center}

\begin{figure}[!ht]
\centering
\includegraphics[width=13cm, height = 7cm]{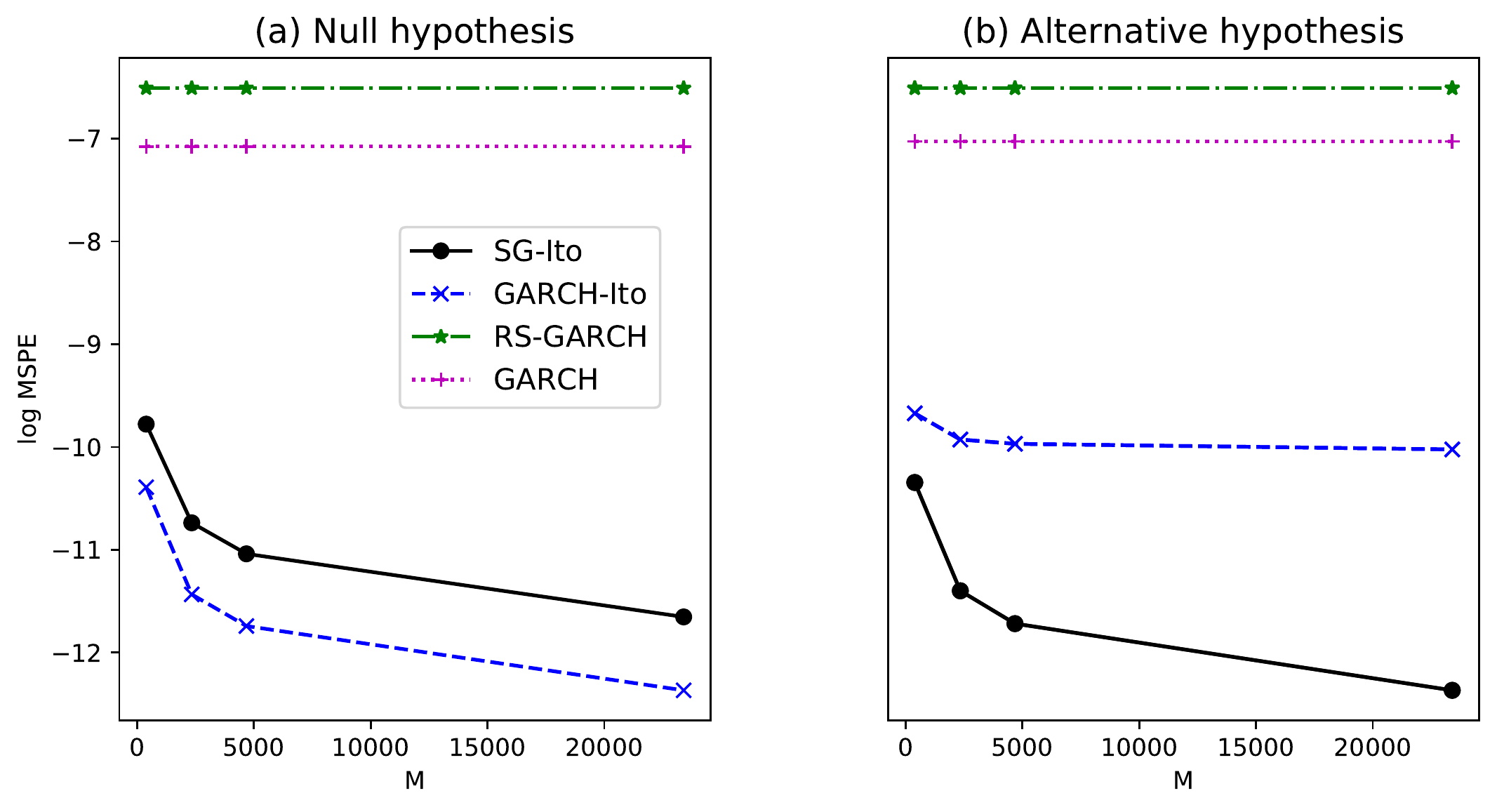}
\caption {One-day ahead out-of-sample volatility prediction error}
\begin{flushleft}
\small{\textit{Notes.}  This figure illustrates one-day ahead out-of-sample volatility prediction error (MSPE) of volatility models against $M$ under the null and alternative hypothesis, with 500-day estimation window and prediction periods. Note that we took the log transformation of MSPEs.}
\end{flushleft}
\label{Figure-3}
\end{figure}

To investigate the advantage of considering state heterogeneity in the model, we compared the prediction performance of the SG-It\^o model with that of the existing volatility models including GARCH(1,1), RS-GARCH(1,1), and unified GARCH-It\^o models. 
Conditional daily volatility processes of the GARCH and unified GARCH-It\^o models are presented as follows:
\begin{align}
    &h_n(\theta^L) = \omega^{L}+\gamma^{L}h_{n-1}(\theta^L)+\beta^{L}\Zna \label{gmodel},\\ 
    % &h_n(\theta^{L}_s) = \omega^{L}_i+\gamma^{L}_i h_{n-1}(\theta^{L}_s)+\beta^{L}_i\Zna  \label{rsgmodel},\\
    &h_n(\theta^g) = \omega^{g*} + \gamma^{g} h_{n-1}(\theta^g) + \beta^{g*}\Zna \label{gitoeq3},
\end{align}
where $\theta^{L} = (\omega^{L}, \gamma^{L}, \beta^{L})$ is the model parameter of the GARCH model, and $\theta^g$ is the model parameter of the unified GARCH-It\^o model described in Appendix \ref{appendix garchito}.
The RS-GARCH(1,1) model is illustrated in Equation \eqref{eq RS-GARCH}.
The QMLE of $\theta^{L}$ and $\theta^{s, L}$ were obtained as in  Equation \eqref{eq lf qml function}, whereas that of the unified GARCH-It\^o parameter was obtained by maximizing Equation \eqref{eq qml function}.
Note that in parameter estimation, the discrete-time models employ daily return square, whereas the continuous-time models employ daily realized volatility estimates. 
To evaluate one-day ahead out-of-sample prediction performances, we calculated the mean squared prediction error (MSPE) of each model as follows:
\begin{align*}
    MSPE =\frac{1}{d}\sum_{n=1}^{d}({RV_n - V_n})^2,
\end{align*}
where $V_n$ is a fitted variance generated from each volatility model, $d$ is the length of prediction window, and the length of both estimation and prediction windows are set to $500$.
Figure \ref{Figure-3} draws the log MSPEs of the GARCH, RS-GARCH, unified GARCH-It\^o, and SG-It\^o models under the null and alternative hypotheses.
In the comparison of the discrete-time models (i.e., GARCH and RS-GARCH) and continuous-time models (i.e., unified GARCH-It\^o and SG-It\^o), we find that the continuous-time models perform better. 
This may be because the discrete-time models need relatively long time periods to obtain consistent estimators, whereas the continuous-time models can estimate the model parameters well in short time period.
This improvement stems from the efficiency of the realized volatility estimator.
%Decrease in the prediction error with the data freqneucy $M$ also supports our claim.
In the comparison of the unified GARCH-It\^o and SG-It\^o models, under the null hypothesis, the unified GARCH-It\^o model performs better.
This is because under the null hypothesis, the unified GARCH-It\^o model is true, so the complexity of the SG-It\^o model brings the inefficiency of parameter estimation. 
On the other hand, the SG-It\^o model outperforms under the alternative hypothesis.
This may be because  the SG-It\^o model can deal with the state heterogeneity in volatility dynamics, whereas the unified GARCH-It\^o model cannot.
% \textbf{Moreover, SG-It\^o (pred) in Panel (b) denotes the case of unrevealed $s_n$ and given $s_{n-1}$.
% % For the state transition probability, we calculate the portion of transitions in every our-of-sample prediction.
% The result suggests that the SG-It\^o model provide superior forecasts compared to the others, even when the future state is not observed.}

%In particular, MSPE of the GARCH (0.0212 and 0.0223 under the null and alternative hypothesis, respectively) and the RS-GARCH (0.0266 and 0.0270) model forecasts are significantly greater than that of the GARCH-It\^o and SG-It\^o models.
%It implies the superiority of continuous-time models, in specific, use of the realized volatility estimator.
%Moreover, Figure \ref{Figure-3} illustrates that GARCH-It\^o model performs better under the null hypothesis due to the complexity of the SG-It\^o model.
%In contrast, the SG-It\^o model outperforms under the alternative hypothesis, implying that the state heterogeneity model would outperform in the case that the true volatility has a state heterogeneity process. 

\begin{center}
    \textbf{[Figure \ref{Figure-3} inserted about here]}
\end{center}

\begin{figure}[!ht]
\centering
\includegraphics[width=1\textwidth]{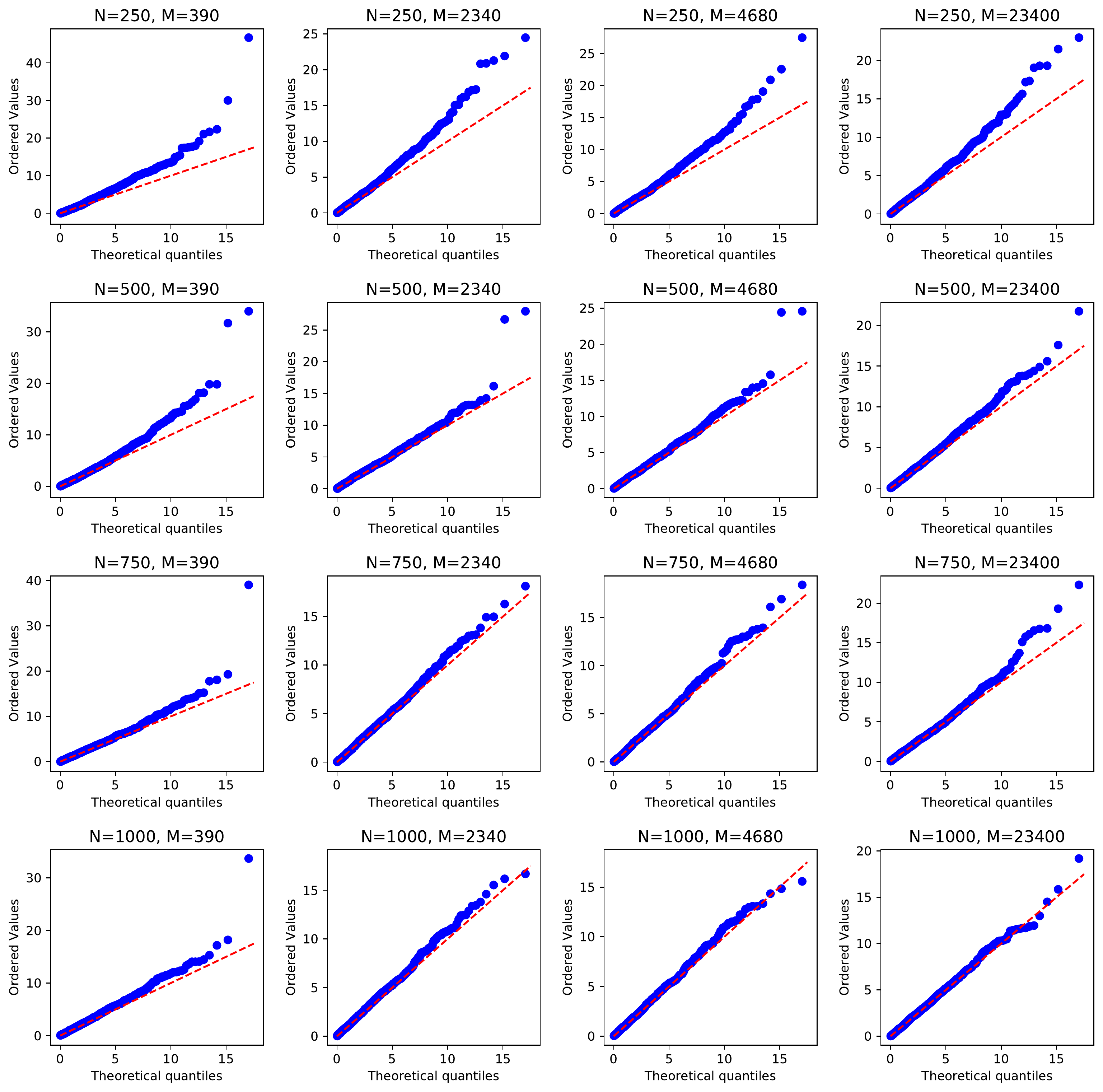}
\caption {$\chi^2$ quantile-quantile plots of the Wald-type statistic}
\begin{flushleft}
\small{\textit{Notes.} This figure illustrates $\chi^2$ quantile-quantile plots of the Wald-type statistic under the null hypothesis for $N = 250, 500, 750, 1000$, and $M = 390, 2340, 4680, 23400$. The real line denotes the best linear fitted line which illustrates perfect $\chi^2$ distribution.}
\end{flushleft}
\label{Figure-4}
\end{figure}

To check the performance of the Wald-type test statistic $T_{N,M}$ developed in Section \ref{sec4}, we investigated the asymptotic convergence of the statistic and conducted size $\alpha$ tests.
Figure \ref{Figure-4} reports $\chi^2$ quantile-quantile plots of the Wald-type statistic $T_{N,M}$ by varying $N,M$ under the null hypothesis. 
The real line in the figures denotes the best linear fitted line that illustrates perfect $\chi^2$ distribution.
Figure \ref{Figure-4} shows that the Wald-type statistic $T_{N,M}$ gradually closes to the limiting distribution $\chi^2$ as $N$ and $M$ increase.
Table \ref{simulation table1} reports the rejection rate of hypothesis test for significance levels of 0.1, 0.05, 0.025, 0.01 by varying $N,M$ under the null and alternative hypotheses.
In Table \ref{simulation table1}, we find that under the null hypothesis, the type I error becomes closer to the suggested significance level $\alpha$ as $N$ and $M$ increase.
That is, the proposed test procedure satisfies size $\alpha$ tests asymptotically. 
Under the alternative hypothesis, the power becomes closer to one as $N$ and $M$ increase. 
These results support the theoretical findings in Section \ref{sec4}. %and support that $(N,M) = (1000, 23400)$ is large enough to test our hypothesis.

% The null and alternative hypothesis are presented in Section \ref{sec4}.
% First and final four columns are for null and alternative hypothesis, respectively.

\begin{center}
    \textbf{[Table \ref{simulation table1} inserted about here]}
\end{center}

\begin{center}
    \textbf{[Figure \ref{Figure-4} inserted about here]}
\end{center}

\subsection{Empirical studies} \label{sec empirical study}

In the empirical study, we examined the volatility process of S\&P 500 index return under the SG-It\^o framework.
We used intraday S\&P 500 index data from 9:30 a.m. to 4:00 p.m., spanning from January 2, 2015, to December 31, 2018 ($N = 998$), provided by Chicago Board of Exchange. 
Before July 23, 2015, data sampling frequency varied from one to three seconds, so the number of intraday data $M_n$ varied from 10,000 to 23,400. 
After July 24, 2015, $M_n$ was fixed to 23,400 except for early closing days. 
We constructed daily pre-averaging realized volatility estimates using intraday index data.

For the SG-It\^o model, the state process plays a prominent role in model specification.
In this empirical study, we considered seven state processes that are known to affect financial volatility and defined models (\romannum{1})--(\romannum{7}) corresponding to each state variable.
The first two models are related to market returns.
These models deal with the negative correlation between financial return and future volatility, which is called the leverage effect \citep{black1976studies, christie1982stochastic, figlewski2000leverage, tauchen1996volume}.
For the market return states, we calculated (\romannum{1}) open-to-close returns for the the previous day market return and (\romannum{2}) close-to-open returns for the overnight return.
%Since the former contains newer information, it is expected to have a greater negative impact on the one-day-ahead market volatility. 
We assigned $s_n = 1$ if (\romannum{1}) the open-to-close return was included in the lowest three deciles and (\romannum{2}) the overnight return was negative, respectively, and $s_n = 0$ otherwise. 
Note that models (\romannum{1}) and (\romannum{2}) incorporate the GJR-GARCH model.
Second, we considered the Chinese stock market information in model (\romannum{3}).
As the second-largest economy in the world, Chinese economy and their stock market may comove with that of the U.S. 
Moreover, the Chinese stock market indices contain information for the non-trading hours in the U.S. stock market.
Thus, we suppose that the Chinese stock market movement affects the U.S. stock market volatility.
We assigned $s_n = 1$ if the Hang Seng index return was included in the lowest three deciles and $s_n = 0$ otherwise. 
Third, we considered the day-of-week seasonality in the financial market, especially on pre- and post-holiday \citep{abraham1994individual, french1980stock,  lakonishok1990weekend, miller1988weekend}. 
%there is some studies about the day-of-week seasonality in financial market, especially on pre- and post-holiday, which is related to the information flow 
Specifically, for models (\romannum{4}) and (\romannum{5}), we constructed pre- and post-holiday indicators using NYSE holiday data and assigned $s_n = 1$ on day (\romannum{4}) before and (\romannum{5}) after NYSE holidays, including weekends, respectively, and $s_n = 0$ otherwise. 
Fourth, we considered trading volume and investor attention.
Previous studies showed that they are positively correlated with financial volatility \citep{andrei2014investor, copeland1976model, jennings1981equilibrium, lamoureux1990heteroskedasticity, lamoureux1994endogenous}.
We measured trading volume and investor attention together with abnormal trading volume $abtv$, calculated by the aggregate market daily dollar volume divided by the sum of recent 20 days dollar volume \citep{barber2007all}.
The large $abtv$ presents the day of high trading volume and high investor attention.
For the model (\romannum{6}), we assigned $s_n = 1$ if the day with $abtv$ was greater than average, and $s_n = 0$ otherwise.
Finally, we adopted an illiquidity measure to proxy the bid-ask based aggregate market illiquidity \citep{chen2018micro, wang2000trading}.
The \citet[CS]{corwin2012simple} measure gauges the illiquidity of individual stocks based on daily high-low spread as follows:
\begin{align*}
    &cs = \frac{2(e^\delta-1)}{1+e^\delta},\\
    &\delta = \frac{\sqrt{2\tau}-\sqrt{\tau}}{3-2\sqrt{2}} - \sqrt{\frac{\rho}{3-2\sqrt{2}}},\ \tau = \left[\log\left(\frac{H_{t-1}}{L_{t-1}}\right)\right]^2+\left[\log\left(\frac{H_{t}}{L_{t}}\right)\right]^2,\ \rho = \left[\log\left(\frac{H_{t-1,t}}{L_{t-1,t}}\right)\right]^2,
\end{align*}
where $H_{t-1,t}$ and $L_{t-1,t}$ are high and low price over days $t-1$ and $t$, respectively.
% \cite{corwin2012simple} use $t+1$ price to solve quadratic equation, but we use $t-1$ price instead because  $H_{t,t+1}$ and $L_{t,t+1}$ are not available on day $t$. 
For model (\romannum{7}), we calculated firm-specific CS measures and value-weighted them to construct the aggregate market illiquidity measure $vwcs$.
We assigned $s_n = 1$ if $vwcs$ was in the highest three deciles, which denoted an illiquid day, and $s_n = 0$ otherwise. 
%We verify that the results are robust by criteria.

Table \ref{parameter table} reports the SG-It\^o model parameter estimation and hypothesis test results. %based on the realized volatility estimates. 
%In the SG-It\^o model, $\omega$ is a scale parameter and $\gamma$ and $\beta$ determine the impact of previous volatility and return shocks on the present volatility, respectively, and 
%All parameters except for $\wb$ are significant at the 1\% level.
The parameter estimates provide some interesting features of volatility processes. 
For example, $\wb$ and $\gb$ of the models (\romannum{1}), (\romannum{2}), and (\romannum{3}) are significantly higher than $\wa$ and $\ga$, respectively, which means that the volatility is generally greater after negative return shock and their clustering become strengthened.
In particular, the greater $\bb$ of the model (\romannum{2}) (0.212) than the model (\romannum{1}) (0.136) may suggest that the market volatility is more sensitive to overnight shocks than the previous day market returns. 
Parameter estimates of model (\romannum{6}) suggest that the impacts of the previous volatility and return shocks on the present volatility increase with heavy tradings.

\begin{center}
    \textbf{[Table \ref{parameter table} inserted about here]}
\end{center}

The results of hypothesis testing suggest that the null hypothesis $H_0: \{\wa = \wb,\ \ga = \gb,\ \ba = \bb\}$ is rejected at the 1\% level for models (\romannum{1})--(\romannum{4}), (\romannum{6}), and (\romannum{7}). 
This implies that the volatility process is distinguished from the homogeneous volatility process when (\romannum{1}) previous day open-to-close return is significantly low, (\romannum{2}) overnight return is negative, (\romannum{3}) Hang Seng index return is significantly low, (\romannum{4}) investors prepare for upcoming holidays, (\romannum{6}) aggregate trading volume is abnormally high, and (\romannum{7}) the market is illiquid.
These results are in line with existing studies. 
For example, \cite{braun1995good}, \cite{carr2017leverage}, and \cite{kim1994alternative} demonstrated that market volatility is significantly increased by negative return shocks.
\cite{ahoniemi2013overnight}, \cite{ahoniemi2016overnight}, and \cite{tsiakas2008overnight} reveal that the overnight information significantly affect the stock market dynamics and help forecast asset volatility.
In particular, the Asian stock markets possibly reflect overnight information of the U.S. stock market because of their time lag \citep{taylor2007note}.
We provide the evidence of close relationship between the U.S. and Chinese stock markets.
\cite{gallant1992stock}, \cite{kambouroudis2016does}, and \cite{karpoff1987relation} showed that aggregate trading volume is positively related to future market volatility. 
The existence of day-of-week and holiday effect are remain up for debate. 
\cite{berument2001day} and \cite{kiymaz2003day} showed the existence of day-of-week effect on market volatility, whereas \cite{birru2018day} claimed that the effect has disappeared on an aggregate level. 
The hypothesis testing results for models (\romannum{4}) and (\romannum{5}) suggest that the pre-holiday effect on the market volatility process may exist, whereas the post-holiday effect has disappeared.

Table \ref{gparameter table} shows the integrated form of the SG-It\^o model parameter estimates described in Theorem \ref{thm1}(b).
Table \ref{all gparameter table} presents parameter estimates of GARCH(1,1), unified GARCH-It\^o (i.e., $\omega^{g*}$, $\gamma^{g*}$, $\beta^{g*}$), and RS-GARCH(1,1) models.
The integrated form of the SG-It\^o model parameters can be interpreted similarly to the RS-GARCH(1,1) model parameters. 
For example, the large $\bbh$ and $\bdh$ of the model (\romannum{2}) in Table \ref{gparameter table} may suggest that daily integrated volatility is significantly affected by market return after negative overnight return shocks.
This is in line with the large $\beta^L_2$ of the RS-GARCH model (\romannum{2}) in Table \ref{all gparameter table}.

\begin{center}
    \textbf{[Table \ref{gparameter table} inserted about here]}
\end{center}

\begin{center}
    \textbf{[Table \ref{all gparameter table} inserted about here]}
\end{center}

%The parameter estimation based on the realized volatility estimates and corresponding hypothesis test results suggest the existence of state heterogeneity in financial volatility. 
 
To investigate the efficiency of adopting the high-frequency data, we estimated the SG-It\^o model parameters and Wald-type statistics using low-frequency data only.
%However, due to its inefficiency, low-frequency data may not produce robust parameter estimates, and consequently, fail to detect state heterogeneity.
%To verify this, we estimate the SG-It\^o model parameters and Wald-type statistics using low-frequency data only.
We estimated parameter estimates $\th^L$ with low-frequency data using the Equation \eqref{eq lf qml function}.
Then we can calculate the Wald-type statistic for $\th^L$, as follows:
\begin{align*}
T_{N}= N(R\th^L-r)^T(R\hat{W}(\hat{V}^{L})^{-1}\hat{W}R^T)^{-1}(R\th^L-r)\xrightarrow{\ d\ }\chi^2(v),
\end{align*}
where
\begin{align*}
    \hat{V}^{L}=\frac{1}{4N}\sumn\left[\dht\dhtt\Big|_{\theta = \th}\ h_n(\th)^{-4} \left(\zeta_n^2-h_n(\th)\right)^2\right].
\end{align*}
Table \ref{parameter table lf} reports SG-It\^o model parameter estimation and hypothesis test results based on low-frequency data only.
We find that standard errors of parameters are significantly increased compared with the results in Table \ref{parameter table}, and, accordingly, most of the $\omega$s and $\beta$s are not significant at the 1\% level anymore.
Moreover, the Wald-type test fails to detect the state heterogeneity in models (\romannum{2})-(\romannum{4}), and the significance of rejection has reduced for models (\romannum{1}) and (\romannum{6}) as well.
These results may imply that the relatively short period of low-frequency data may not contain sufficient information and fail to capture low-frequency volatility dynamics.
From the results, we can conclude that the use of high-frequency data helps to analyze low-frequency dynamics for relatively short-time-period data, so it would be more robust to the structural break issue.
These findings support our hypotheses of existence of state heterogeneity and efficiency of using high-frequency data to examine low-frequency market dynamics.

\begin{center}
    \textbf{[Table \ref{parameter table lf} inserted about here]}
\end{center}

Foregoing results indicate the existence of state heterogeneity in the volatility process for well-known state variables.
This may imply that the volatility model considering state heterogeneity would produce better volatility estimates and forecasts.
% This may imply that the information from state- and time-domains are exclusive, and therefore, the model which takes both information into account would produce better volatility estimates and forecasts.
% True models tend to have superior prediction performance and it is more evident for the volatility models because of the clustering effect. 
To compare the prediction performance of representative volatility models, we implemented the one-day-ahead out-of-sample volatility prediction and calculated mean absolute percentage error (MAPE), presented as follows:
\begin{align*}
    MAPE = \frac{100}{d}\sum_{n=1}^{d}\left|\frac{RV_n - V_n}{RV_n}\right|,
\end{align*}
where $d$ is the length of prediction window and $V_n$ is a fitted variance generated from each volatility model.
The estimation window is 750 days with the prediction period spanning from December 22, 2017, to December 31, 2018 (248 days).
The benchmarks are unified GARCH-It\^o, RS-GARCH(1,1), and GARCH(1,1) models and the model specifications are presented in Section \ref{sec simulations}.
We also consider heterogeneous auto-regressive (HAR) model of \cite{corsi2009simple} as an additional benchmark.
On the one hand, state variable $s_n$ is available at the beginning of day $n$ for models (\romannum{1}) previous day open-to-close return, (\romannum{2}) overnight return, (\romannum{3}) Hang Seng index return, (\romannum{4}) pre-holiday, and (\romannum{5}) post-holiday.
On the other hand, $s_n$ is not observed until the end of day $n$ for the models (\romannum{6}) abnormal trading volume and (\romannum{7}) market illiquidity, so we have to utilize a state transition probability as in Proposition \ref{prop1}.
To obtain state transition probability, we simply assumed time-persistent state transition probability and calculated the portion of transition from state $j$ to $i$ for $p_{ij}$.
The estimation of state transition probability significantly affects prediction performance, but we leave a more elaborate probability inference for further research.
Table \ref{prediction table} reports out-of-sample prediction results measured by MAPEs.
The results suggest that continuous-time models (SG-It\^o and unified GARCH-It\^o) performs better than discrete-time models (GARCH and RS-GARCH) and the HAR model and that state heterogeneity models (SG-It\^o and RS-GARCH) are superior to state-homogeneous models (GARCH and unified GARCH-It\^o) in general.
Thus, the continuous and state heterogeneous SG-It\^o model shows outstanding performance compared to the others.
In particular, the MAPE improvement of the SG-It\^o model is the greatest in model (\romannum{2}), in which the state heterogeneity was the greatest, whereas the improvement seems insignificant in model (\romannum{5}), in which state heterogeneity was not detected.
For models (\romannum{6}) and (\romannum{7}), although we employed the simple procedure to estimate transition probability, the prediction performance of the SG-It\^o model is similar to or even better than that of the benchmark models.

\begin{center}
    \textbf{[Table \ref{prediction table} inserted about here]}
\end{center}

\section{Conclusions} \label{sec6}

State heterogeneity in financial volatility has widely been discussed as a representative market characteristic. 
% integration of time- and state-domain information provides better statistical fit and prediction performance.
This study hypothesizes that there exists state heterogeneity in financial volatility and use of high-frequency data facilitates analyzing it.
To test the hypothesis, we proposed a novel volatility model whose instantaneous volatility has a continuous-time process and that evolves depending on the discrete state process.
% This study contributes to the field of finance by introducing an advanced approach to deal with state heterogeneity in volatility and examine low-frequency market dynamics using informative high-frequency financial data.
Through the model, this study provides a  mathematical background to apply high-quality realized volatility estimators to the study of discrete-time state heterogeneity volatility frameworks. 
% which is an effective framework to unify the information from continuous-time and discrete-time state process in volatility analysis.
% Moreover, this model allows to analyze low-frequency market dynamics with high-frequency modeling 
Along with the model, we construct a Wald-type hypothesis testing procedure to test our hypothesis.
Through hypothesis testing, we verify the existence of leverage, investor attention, market illiquidity, stock market comovement, and post-holiday effect in S\&P 500 index volatility. 
The statistical test based on low-frequency data only, however, does not catch these effects well.

In this paper, our focus is to test the given exogenous state.
However, in practice, how to define the state process is an important but difficult question.
Fortunately, the proposed SG-It\^o diffusion process is not affected by the state process, so it is easy to incorporate any state process in the SG-It\^o process structure. 
Thus, studying state processes based on the high-frequency financial data is a promising direction for future research.

% using realized volatility estimates, while low-frequency data hardly detect the state heterogeneity.

% The superior prediction performance of SG-It\^o model supports the advantage of considering state heterogeneity in financial volatility using high-frequency data.

% Appendix here
% Options are (1) APPENDIX (with or without general title) or
%             (2) APPENDICES (if it has more than one unrelated sections)
% Outcomment the appropriate case if necessary
%
% \begin{APPENDIX}{<Title of the Appendix>}
% \end{APPENDIX}
%
%   or
%
% \begin{APPENDICES}
% \section{<Title of Section A>}
% \section{<Title of Section B>}
% etc
% \end{APPENDICES}

\clearpage
\appendix
\addcontentsline{toc}{section}{Appendix}
\section{Appendix} \label{sec appendix}
\subsection{Instantaneous volatility}

Under the SG-It\^o framework, the instantaneous volatility at integer time point $n$ can be presented as the linear function of $\sigma_{n-1}^2$ and daily return square as follows:
\begin{align*}
    &\sigma_n^2 = \yaa\sigma_{1,n}^2+\yba\sigma_{2,n}^2\\
    &\quad=\yaa(\wa+\ba\Za)+\yba(\wb+\bb\Zb)+(\yaa\ga+\yba\gb)\sna.
\end{align*}
Then  we can express instantaneous volatility at integer time point as the infinite sum of $Z_{i,n}$'s using a recursive relationship. 
Let $D_n(k)=[(1-s_{n+1-k})\ga+s_{n+1-k}\gb]D_n(k-1)$ and $D_n(0)=1$. 
Then $D_n(k)=\prod_{i=0}^{k-1}[ (1-s_{n-i})\ga+ s_{n-i}\gb]$ and we have
\begin{align}\label{eq sig cumsum}
\begin{split}
    \sn&=D_n(0)\{\yaa(\wa+\ba\Za)+\yba(\wb+\bb\Zb)\}+D_n(1)\sna\\
    &=D_n(0)\{\yaa(\wa+\ba\Za)+\yba(\wb+\bb\Zb)\}\\
    &\quad+D_n(1)\{\yab(\wa+\ba\Zaa)+\ybb(\wb+\bb\Zba)\}+D_n(2)\snb\\
    &=\sum_{i=0}^{k-1}D_n(i)\{(1-s_{n-i})(\wa+\ba Z_{1,n-i}^2)+s_{n-i}(\wb+\bb Z_{2,n-i}^2)\}+D_n(k)\si_{n-k}^2\\
    &=\sum_{i=0}^{\infty}D_n(i)\{(1-s_{n-i})(\wa+\ba Z_{1,n-i}^2)+s_{n-i}(\wb+\bb Z_{2,n-i}^2)\} \text{ a.s.}
\end{split}
\end{align}
Note that $D_n(k) \le (\guh)^k$, $\sn$ satisfies following inequality:
\begin{align*}
    \sn\le\sum_{i=0}^{\infty}(\guh)^i (\wuh+\buh Z_{n-i}^2)
    =\frac{\wuh}{1-\guh}+\buh\sum_{i=0}^{\infty}(\guh)^{i} Z_{n-i}^2.
\end{align*}
Then, by Assumption \ref{ass1}(d), we can easily show the existence of the infinite sum in \eqref{eq sig cumsum}.

\subsection{Connection with the GARCH-It\^o model} \label{appendix garchito}
We can show that when the states are homogeneous ($s_n = s_{n-1} = \cdots = s_1$), the SG-It\^o model returns to the unified GARCH-It\^o model  \citep{kim2016}. 
That is, the unified GARCH-It\^o model is a special example of the SG-It\^o model. 
The unified GARCH-It\^o model can be presented as follows:
\begin{align}
    &dX_t = \mu dt + \sigma_t dB_t, \label{eq gitoeq1}\\
    &\sigma_t^2=\sigma_{n-1}^2+(t-n+1)\{\omega^g+(\gamma^g-1)\sigma_{n-1}^2\}+\beta^g \left(\int_{n-1}^{t} \sigma_s dB_s \right)^2, \label{eq gitoeq2}
\end{align} 
where $\theta^g = (\omega^g, \beta^g, \gamma^g)$ are model parameters. 
Let us assume $s_n = 1$ for all $n \in \mathbb{N}$. 
Then, by Equation \eqref{eq sig cumsum}, the instantaneous volatility under the SG-It\^o model can be presented as follows:
\begin{align*}
    \sn&=\sum_{i=0}^{\infty}\ga^{i}(\wa+\ba Z_{n-i}^2)
    =\frac{\wa}{1-\ga}+\ba\sum_{i=0}^{\infty}\ga^{i}Z_{n-i}^2.
\end{align*}
For $\hn$, we have
\begin{align*}
    \hn&=\wah+\gah\hna+\bah\Zna\\
    &=\omega^{g*}+\gamma^g h_{n-1}(\theta)+\beta^{g*} Z_{n-1}^2,
\end{align*}
where $\omega^{g*} = \omega^g(\beta^g)^{-1}(e^{\beta^g}-1)$ and $\beta^{g*} = (\beta^g)^{-1}(\gamma^g-1)(e^{\beta^g}-1-\beta^g)+e^{\beta^g}-1$.

\subsection {Proof of Theorem \ref{thm1}}
\proof{Proof of Theorem \ref{thm1}}
Consider $(a)$.
By It\^o's lemma, we have
\begin{align*}
    R_1(k)&=\int_{n-1}^n \frac{(n-t)^k}{k!}\sigma_{1,t}^2 dt\\
    &=\int_{n-1}^n \frac{(n-t)^k}{k!}[\sna+(t-n+1)\{\wa+(\ga-1)\sna\}]dt \\
    &\quad+\ba\int_{n-1}^{n}\frac{(n-t)^k}{k!}\left(\int_{n-1}^{t} \sigma_{1,s} dB_{s} \right)^2dt \\
    &=\frac{1}{(k+2)!}[\wa+(\ga+k+1)\sna]\\
    &\quad+2\ba\int_{n-1}^{n}\frac{(n-t)^{k+1}}{(k+1)!} \left(\int_{n-1}^{t} \sigma_{1,s} dB_{s}\right)\sigma_{1,t}dB_{t}+\ba R_1(k+1).
\end{align*}
Then we have
\begin{align*}
    \nint \si_{1,t}^2dt=R_1(0)=\ba^{-2}(e^{\ba}-1-\ba)\wa + [(\ga-1)\ba^{-2}(e^{\ba}-1-\ba)+\ba^{-1}(e^{\ba}-1)]\sna+\xi_{1,n},
\end{align*}
where $\xi_{1,n}=2\nint(e^{(n-t)\ba}-1)\int_{n-1}^{t}\si_{1,s}dB_{s}\si_{1,t}dB_{t}$.
We can calculate $\nint \si_{2,t}^2dt$ in the same way. 
Then  integrated volatility under the SG-It\^o framework can be expressed as a function of \{$\mathcal{F}^{x,L}_{n-1},\mathcal{F}^{s}_{n-1}$\}-adapted process and the martingale difference as follows:
\begin{align*}
    &\nint \si_{i,t}^2dt=h_{i,n}(\theta)+\xi_{i,n,} \\ 
    &E\left[\nint \si_{i,t}^2dt\bfilta\right] = h_{i,n}(\theta)=\Hic+\Hib\sna \text{ a.s.}
\end{align*}

Consider $(b)$. 
By the result of $(a)$, we have
\begin{align*}
      E\left[ \int_{n-1}^n \sigma_t dt \bfilta\right]  &= h_n(\theta)=\yaa\ha+\yba\hb\quad\\ 
    & =\yaa\left(\Hac+\Hab\sna\right)+\yba\left(\Hbc+\Hbb\sna\right)\\
    &=\aHc+\aHb\sna,
\end{align*} 
where $\aHc = \yaa\Hac + \yba\Hbc$ and $\aHb = \yaa\Hab + \yba\Hbb$.
By Equation \eqref{eq sig cumsum}, we have
\begin{align*}
      &h_n(\theta) =\aHc+\aHb\sna \\
      & = \aHc+\aHb\sum_{i=0}^{\infty}D_{n-1}(i)\{(1-s_{n-1-i})(\wa+\ba Z_{1,n-1-i}^2)+s_{n-i}(\wb+\bb Z_{2,n-1-i}^2)\} \\
      & = \aHc+\aHb\sum_{i=1}^{\infty}D_{n-1} (i)\{(1-s_{n-1-i})(\wa+\ba Z_{1,n-1-i}^2)+s_{n-i}(\wb+\bb Z_{2,n-1-i}^2)\} \\	
      &\quad +\aHb\{(1-s_{n-1})(\wa+\ba Z_{1,n-1}^2)+s_{n-1}(\wb+\bb Z_{2,n-1}^2)\} \\
      & = \aHc+ \aHb D_{n-1} (1)  \sigma_{n-2}^2   +\aHb\{(1-s_{n-1})(\wa+\ba Z_{1,n-1}^2)+s_{n-1}(\wb+\bb Z_{2,n-1}^2)\}.    
\end{align*}
Thus,  we have
\begin{align}\label{eq-hn}
    \hn &=\aw+\agij\hna+\ab\Zna, 
\end{align}
where $\aw = \aHc +\aHb\{(1-s_{n-1})\wa +s_{n-1} \wb \} - \agij H_{c,n-1}^\omega (\theta) $, $ \agij = \frac{D_{n-1} (1) H_{\beta,n} ^\omega (\theta)}{ H_{\beta,n-1} ^\omega (\theta) }$, $\ab= \aHb\{(1-s_{n-1})\ba +s_{n-1}\bb \} $. 
Then we can easily show
\begin{align*}
\begin{split} 
    \hn&=\yaya(\wah+\bah\hna+\gah\Zna)+\yayb(\wbh+\bbh\hna+\gbh\Zna)\\
    &\quad+\ybya(\wch+\bch\hna+\gch\Zna)+\ybyb(\wdh+\bdh\hna+\gdh\Zna).
\end{split}
\end{align*}
\endproof

\subsection{Proof of asymptotic theories} \label{appendix asymptotic theory}
This section provides proofs of asymptotic theories presented in Section \ref{sec3}. 
First  Lemma \ref{lemma initial value} shows that the impact of the initial value is asymptotically negligible by showing that the impact of initial value on $\hn$ is exponentially decaying. 
Accordingly, the difference between the quasi-likelihood functions with true and arbitrary value decays faster than $O_p(N^{-1})$.

\subsubsection{Initial value}
\begin{lemma} Under Assumption \ref{ass1}(a), we have for any $\vartheta=O_p(1)$ and $\ninN$, $|h_n(\theta_0,\sigma_0^2)-h_n(\theta_0,\vartheta)|=O_p((\guh)^{n-1})$. \label{lemma initial value}
\end{lemma}
\proof {Proof.} 
%We have
%\begin{align*}
%    E\left[Z_n^2\sfilta\right]& =E[\hon\sfilta].
%\end{align*} 
Simple algebraic manipulations provide   
\begin{align*}
    h_n(\theta_0,\sigma_0^2)-h_n(\theta_0,\vartheta)&=\prodp\agp(h_1(\to, \sigma_0^2) - h_1(\to, \vartheta))\\
    &\le (\guh)^{n-1}\aHb[1](\sigma_0^2-\vartheta)=O_p((\guh)^{n-1}).
\end{align*}
Thus, as $N\rightarrow\infty$, the difference between $\sigma_0^2$ and $\vartheta$ become negligible.
\endproof

\subsubsection{Proof of Theorem \ref{thm2}}
For given \{$s_{n}$\}, we define the log likelihood functions and their derivatives as follows:
\begin{align*}
    &\Lhnm=-\frac{1}{2N}\sumn \left[\log(\hn)+\frac{RV_n}{\hn}\right]= -\frac{1}{2N}\sumn \lhn,\\
    &\phnm= \frac{\partial\Lhnm}{\partial\theta},\quad \th = \mathop{\mathrm{argmax}}_{\tinT} \Lhnm,\\
    &\Lhn=-\frac{1}{2N}\sumn \left[\log(\hn)+\frac{\isign}{\hn}\right],
    \quad\phn= \frac{\partial\Lhn}{\partial\theta},\\
    &\Ln=-\frac{1}{2N}\sumn \left[\log(\hn)+\frac{\hon}{\hn}\right],
    \quad\pn= \frac{\partial\Ln}{\partial\theta}.
\end{align*}
We denote derivatives of function $g$ at $x^*$ by $\frac{\partial g(x^*)}{\partial x} = \frac{\partial g(x)}{\partial x}\Big|_{x = x^*}$. 
Note that in Assumption \ref{ass1}(a), we defined upper and lower bounds of $\theta$ and $\theta^h$.
\begin{lemma} Under Assumption \ref{ass1}(a), we have
\begin{itemize}
    \item[(a)]$\supn E[\Zn]\le\frac{\wuh}{1-\buh-\guh}+E[h_1(\to)]<\infty,$\ 
    and $\supn E[\supt\hn]<\infty$.
    \item[(b)]$\xi_{i,n}=\beta_{0,i}\nint e^{(n-t)\beta_{0,i}}(Z_{i,t}-Z_{i,n-1})^2-\nint e^{(n-t)\beta_{0,i}}\sigma^2_{i,t}dt$ a.s. for any $\ninN$ and $i\in\{1,2\}$.
    \item[(c)]There exists a neighborhood $B(\to)$ of $\to$ such that for any $p\ge 1$, $\supn \lpnorm{\supB \frac{\hon}{\hn}}<\infty$ and $B(\to)\subset\Theta$.
    \item[(d)]For any $\ninN$, $\supn\lpnorm{\supB\frac{1}{\hn}\dhj}\le C$, $\supn\lpnorm{\supB\frac{1}{\hn}\ddhjk}\le C$, and $\supn\lpnorm{\supB\frac{1}{\hn}\dddhjkl}\le C$ for any $j,k,l\in\{1,2,3,4,5,6\}$, where $\theta=(\theta_1,\theta_2,\theta_3,\theta_4,\theta_5,\theta_6)=(\wa,\wb,\ga,\gb,\ba,\bb)$.
\end{itemize} \label{lem bounded}
\end{lemma} 

\proof{Proof.}
For $(a)$,  by Equation \eqref{eq-hn},  we can express daily integrated volatility as infinite sum of $Z_{i,n}$'s as follows:
\begin{align*}
    \hn&=\aw+\agij\hna+\ab\Zna\\
    &=\sump(\awp+\abp\Znp)\prodq\agq+\prodp\agp h_1(\theta).
\end{align*}
%$In the proof of Lemma \ref{lemma initial value}, we showed that  
%$E\left[Z_n^2\sfilta\right] = E\left[\nint\si_t^2dt\sfilta\right]=E[h_{n}(\to)\sfilta]$, 
%so it is obvious that $E[Z_n^2]=E\left[\isign\right]=E[\hon]$. 
Using iterative relationship and $\agij+\ab < 1$, we can show that
\begin{align*}
   E[ Z_n^2] &=  E[\hon] =\aw+\agij E[h_{n-1}(\theta_0)]+\ab E[\Zna]\\
    &=\aw+(\agij+\ab) E[h_{n-1}(\to)]\\
    &\le\wuh\frac{1-(\buh+\guh)^{n-1}}{1-(\buh+\guh)}+(\guh)^{n-1}E[h_1(\to)]\\
    &\le\frac{\wuh}{1-\buh-\guh}+E[h_1(\to)]<\infty.
\end{align*}
Then we can easily show $\supn E[\supt\hn]<\infty$.

For $(b)$, let $f(t,Z_{i,t})=(e^{(n-t)\beta_{0,i}}-1)(Z_{i,t}-Z_{i,n-1})^2$. 
Then, by It\^o's lemma, we have
\begin{align*}
    &df(t,Z_{i,t})=\left[-\beta_{0,i}e^{(n-t)\beta_{0,i}}(Z_{i,t}-Z_{i,n-1})^2+(e^{(n-t)\beta_{0,i}}-1)
    \sigma_{i,t}^2\right]dt \\
    &\qquad \qquad \qquad +2(e^{(n-t)\beta_{0,i}}-1)(Z_{i,t}-Z_{i,n-1})dZ_{i,t},\\
    &f(n,Z_{i,n})=0=\nint\left[-\beta_{0,i}e^{(n-t)\beta_{0,i}}(Z_{i,t}-Z_{i,n-1})^2+(e^{(n-t)\beta_{0,i}}-1)\sigma_{i,t}^2\right]dt+\xi_{i,n}.
\end{align*}

Consider $(c)$.
 For any $\delta>0$, there exists a neighborhood $B(\to)\subset\Theta$ such that  $\ago\le(1+\delta)\agij$. 
Using the fact that $x/(1+x)\le x^q$ for all $x \ge 0$ and any $q\in[0,1]$, we have
\begin{align*}
    \frac{\hon}{\hn}&=\frac{\sump(\awop+\abop\Znp)\prodq\agoq+\prodp\agop h_1(\to)}
    {\sump(\awp+\abp\Znp)\prodq\agq+\prodp\agp h_1(\theta)}\\
    &\le\frac{\sump(\awop+\abop\Znp)\prodq\agoq+C}{\aw+\sum_{k=2}^{n-1}(\awp+\abp\Znp\prodq\agq)+C}\\
    &\le\sump\left[\frac{\wuh}{\wlh}(\guh)^{k-1}+\frac{\abop\Znp\prodq\agoq}{\wlh+\abp\Znp\prodq\agq}\right]+C\\
    &=C+\frac{\blh}{\buh}\sump\frac{\abp\Znp\prodq\agq}{\wlh+\abp\Znp\prodq\agq}\left(\frac{\prodq\agoq}{\prodq\agq}\right)\\
    &=C+C\sump \frac{x_k}{1+x_k}\prodq\frac{\agoq}{\agq},\ \\
    &\le C+C\sump x_k^q\prodq\frac{\agoq}{\agq}\\
    &\le C+C\sump(\buh)^q\frac{Z_{n-k}^{2q}}{(\wlh)^q}(\guh)^{q(p-1)}(1+\delta)^{p-1}\\
    &=C+C\sump(\guh)^{q(p-1)}(1+\delta)^{p-1} Z_{n-k}^{2q}\\
    &=C+C\sump{\rho^{p-1}} Z_{n-k}^{2q},
\end{align*}
where $x_k=\frac{\abp\Znp\prodq\agq}{\wlh}$. 
Let $0<\delta<\frac{1-(\guh)^q}{(\guh)^q}$. 
Then, $(1+\delta)<\frac{1}{(\guh)^q}$ and $\rho=(1+\delta)(\guh)^q<1$. 
Taking $q\in[0,1]$ such that $E(Z_{n-k}^{2pq})<\infty$, we have
\begin{align*}
    \lpnorm{\supB\frac{\hon}{\hn}}
    \le C+C\sump\rho^{p-1}\lpnorm{Z_{n-k}^{2q}}<\infty.
\end{align*}
From $\abs{\rho}<1$, we conclude that
\begin{align*}
    \supn\lpnorm{\sup_{\theta\in B(\to)}\frac{\hon}{\hn}}<\infty.
\end{align*}

For (d), we first examine the first derivatives. For $\wa, \wb, \ba,$ and $\bb$, we can show that
\begin{align*}
    \frac{1}{\hn}\dhj \le C \quad \text{a.s. for $j = 1, 2, 3, 4$},
\end{align*}
because $\sn$ is their linear function.

Consider the case that ($\omega_n^w, \beta_n^w, \gamma_n^w$) = ($\wch, \bch, \gch$).
Under Assumption \ref{ass1}(a), we can easily show that
\begin{align*}
    \abs{\frac{\partial\theta^h_j}{\partial\gamma_k}} \le C \quad \text{a.s. for $j = 1, 2, \ldots , 12 $ and $k = 1, 2$}.
\end{align*}
The property that $x/(1+x)\le x^q$ for any $q \in [0, 1]$ and all $x \ge 0$ gives us
\begin{align*}
    \abs{\frac{1}{\hn}\dfunc{\hn}{\ga}}
    &=\hn^{-1}\abs{\sump\dfunc{\wch}{\ga}(\gch)^{k-1}
    + (k-1)(\wch + \bch Z_{n-k}^2)(\gch)^{k-2}\dfunc{\gch}{\ga}}\\
    &\quad + (n-1)(\gch)^{n-2}h_1(\theta)\dfunc{\gch}{\ga}\\
    &\le C\abs{\sump\frac{(\gch)^{k-1}}{(\gch)^{k-1}(\wch + \bch Z_{n-k}^2)}}
    +C\abs{\sump\frac{(k-1)(\gch)^{k-2}(\wch + \bch Z_{n-k}^2)}
    {\wch +(\gch)^{k-2}(\wch + \bch Z_{n-k}^2)}}+C\\
    &\le C \abs{\sum^{n-2}_{k=1}k\rho^{kq}(\wuh+\buh Z_{n-k-1})^q} + C.
\end{align*}
We can choose $q \in [0, 1]$ such that $E(\wuh+\buh Z_{n-k-1})^{qp}<\infty$.
Then, since $\abs{\rho}<1$,
\begin{align*}
    \supn\lpnorm{\frac{1}{\hn}\dfunc{\hn}{\ga}}<C.
\end{align*}
Similarly, we can show the bound for the first derivatives of the $\hn$ and for the second and third derivatives.
\endproof

\begin{lemma} Under Assumption \ref{ass1}(a)--(b) and Assumption \ref{ass2}(a)--(b), we have
\begin{align*}
    \supt\abs{\Lhnm-\Ln}=O_p(M^{-\frac{1}{4}})+o_p(1).
\end{align*}
\end{lemma}
\proof{Proof.}
By the triangular inequality, we have
\begin{align*}
    \abs{\Lhnm-\Ln}\le\abs{\Lhnm-\Lhn}+\abs{\Lhn-\Ln}.
\end{align*}
Note that $\hn^{-1}<\infty$ a.s. 
By Assumption \ref{ass2}(b), we have
\begin{align*}
    &E\left[\supt\abs{\Lhnm-\Lhn}\right]\le C\frac{1}{N}\sumn E\left[\RVsig\right]\le CM^{-\frac{1}{4}}.
\end{align*}
Accordingly, we have
\begin{align*}
    &\supt\abs{\Lhnm-\Lhn}=O_p(M^{-\frac{1}{4}}).
\end{align*}
We can easily show that $\Lhn-\Ln=\frac{1}{2N}\sumn\frac{\xi_n}{\hn}$. 
Because $\hn$ is adapted to $\mathcal{F}_{n-1}$, $\frac{\xi_n}{\hn}$ is also martingale difference. 
Furthermore, $\abs{\frac{\xi_n}{\hn}}$ is uniformly integrable because $\abs{\frac{\xi_n}{\hn}}\le \frac{1}{\wlh}\abs{\xi_n}$. 
Thus, $\abs{\Lhn-\Ln}\longrightarrow 0$ in probability (see Theorem 2.22 in \cite{hall2014martingale}). 

Let $\Gn=\Lhn-\Ln$. 
By mean-value theorem, there exists $\ts$ between $\theta$ and $\theta^{'}$ satisfying
\begin{align*}
    \abs{\Gn-\Gnprime}&=\abs{\frac{1}{2N}\sumn\frac{\xi_n}{h_n^2(\ts)}\dhst(\theta-\theta^{'})}\\
    &\le\frac{1}{2N}\sumn\maxnorm{\sup_{\ts\in\Theta}\frac{\xi_n}{h_n^2(\ts)}\dhst}\maxnorm{(\theta-\theta^{'})}.
\end{align*}
By Lemma \ref{lem bounded}(d), $\ltwonorm{{\partial\hsn\over\partial\theta_k}\frac{1}{\hsn}}\le C$ for every $k\in\{1,2,3,4,5,6\}$. 
Therefore, we have
\begin{align*}
    \lonenorm{\sup_{\ts\in\Theta}\frac{\xi_n}{h_n^2(\ts)}\dhst}\le C\lonenorm{\xi_n}\le C<\infty.
\end{align*}
As a result, $\Gn$ satisfies the weak Lipschitz condtion and uniformly converges to zero by Theorem 3 in \cite{andrews1992generic}.
\endproof

\proof{Proof of Theorem \ref{thm2}.}
First, let us show the existence of the unique maximizer of $\Ln$. 
From the definition of $\Ln$, it is obvious that 
\begin{align*}
    \max_{\tinT}\Ln\le-\frac{1}{2N}\sumn\min_{\theta_n\in\Theta}\left[\log(h_n(\theta_n))+\frac{\hon}{h_n(\theta_n)}\right].
\end{align*}
If $\theta_{0,n}$ is the minimizer of the $n$th summand on right hand side, $\theta_{0,n}$ must satisfy $h_n(\theta_{0,n})=\hon$ for every $\ninN$. 
Therefore, if there exists $\ts\in\Theta$ such that $\hsn=\hon$  for every $\ninN$, $\ts$ would be the maximizer. 
In this manner, $\to$ is one of the candidates of $\ts$. 
We then show that $\ts=\to$ a.s. 

Under the SG-It\^o framework, we have
\begin{align*}
    \hn&=\yaya(\wah+\gah\hna+ \bah\Zna)+\yayb(\wbh+\gbh \hna+ \bbh\Zna)\\
    &\quad+\ybya(\wch+\gch \hna+ \bch\Zna)+\ybyb(\wdh+\gdh\hna+ \bdh\Zna).
\end{align*}
Then $\ts$ and $\to$ satisfy $AP=0$ a.s., where
\begin{align*}
&\ts = (\wa^*, \wb^*, \ga^*, \gb^*, \ba^*, \bb^*),\\
&A=\begin{pmatrix}
    \yayaoa & \cdots & \ybyboa & \yayaoa h_1(\to) & \cdots & \ybyboa h_1(\to) & \yayaoa Z_1^2 & \cdots & \ybyboa Z_1^2\\\yayaob & \cdots & \ybybob & \yayaob h_2(\to) & \cdots & \ybybob h_2(\to) & \yayaob Z_2^2 & \cdots & \ybybob Z_2^2\\\vdots & & \vdots & \vdots & & \vdots & \vdots & & \vdots \\ \yaya & \cdots & \ybyb & \yaya \hn & \cdots & \ybyb \hn & \yaya \Zn & \cdots & \ybyb \Zn
\end{pmatrix},\\
&P^T = (\wahs-\woah\ \wbhs-\wobh\ \wchs-\woch\ \wdhs-\wodh\ \gahs-\goah\ \gbhs-\gobh\\ &\quad\quad\quad\gchs-\goch\ \gdhs-\godh\ \bahs-\boah\ \bbhs-\bobh\ \bchs-\boch\ \bdhs-\bodh).
\end{align*}
Note that $A$ is of full rank because $Z_n^2$ is nondegenerated and $s_n=0$ or $1$. 
Then  $A^TA$ is invertable and $P = 0$ a.s. 
That is, we  have
\begin{align*}
    % \omega_1^{g*}=\wahs=\woah=\omega_{0,1}^g,\\
    % \omega_2^{g*}=\wdhs=\wodh=\omega_{0,2}^g,\\
    % \gamma_1^{g*}=\gahs=\goah=\gamma_{0,1}^g,\\
    % \gamma_2^{g*}=\gdhs=\godh=\gamma_{0,2}^g,\\
    % \beta_1^{g*}=\bahs=\boah=\beta_{0,1}^g,\\
    % \beta_2^{g*}=\bdhs=\bodh=\beta_{0,2}^g.
    \wahs=\woah, \wdhs=\wodh, \gahs=\goah, \gdhs=\godh, \bahs=\boah, \bdhs=\bodh\quad \text{a.s.},
\end{align*}
which implies $\ts=\to$ a.s. 
This also implies that there is a unique maximizer of $\Ln$ because $\wah (\wdh)$ and $\bah (\bdh)$ are strictly increasing function of $\beta_1 (\beta_2)$. 
Then, for any $\epsilon>0$, there is a constant $\nu$ such that
\begin{align*}
    Q_N(\to) - \max_{\tinT:\maxnorm{\theta-\to}>\epsilon}\Ln>\nu \quad \text{a.s.}
\end{align*}
Now, the theorem is the result of Theorem 1 in Xiu (2010). 
\endproof

\subsubsection{Proof of Theorem \ref{thm3}} 
\begin{lemma} \label{lem4} We have following properties under Assumption \ref{ass1}(a), Assumption \ref{ass2}(c), and Lemma \ref{lem bounded}(d):
\begin{itemize}
    \item [(a)] There exists a neighborhood $B(\to)$ of $\to$ such that $\supn\lonenorm{\supB\frac{\partial^3\lhn}{\partial\theta_j\partial\theta_k\partial\theta_l}}<\infty$ for any $j,k,l\in\{1,2,3,4,5,6\}$, where $\theta=(\theta_1,\theta_2,\theta_3,\theta_4,\theta_5,\theta_6)=(\wa,\wb,\ga,\gb,\ba,\bb)$.
    \item [(b)] $-\dpon$ is a positive definite matrix for $N\ge 3$
\end{itemize}
\end{lemma}
\proof{Proof.}
Consider $(a)$.
 By Assumption \ref{ass2}(c), we have
\begin{align*}
    &E[RV_n\sfilta]\le CE\left[\isign\bfilta\right]+C=C\hon+C\ a.s.
\end{align*} Then, by Lemma \ref{lem bounded}(c) and (d), we have
\begin{align*}
    &E\left[\supB\abs{\frac{RV_n}{\hn}\dddhjkl}\right]\\
    &\quad\le CE\left[\supB\frac{\hon}{\hn}\abs{\frac{1}{\hn}\dddhjkl}\right]+C\\
    &\quad\le C\ltwonorm{\supB\frac{\hon}{\hn}}\ltwonorm{\supB\abs{\frac{1}{\hn}\dddhjkl}}+C\le C<\infty.
\end{align*}
We can similarly bound remaining terms.

Consider $(b)$.
 Let 
 $$h_{\theta,n}=\dhot\hon^{-1}=\hon^{-1}\begin{pmatrix} \frac{\partial\hon}{\partial\omega_1} & \frac{\partial\hon}{\partial\omega_2} & \frac{\partial\hon}{\partial\gamma_1} & \frac{\partial\hon}{\partial\gamma_2} & \frac{\partial\hon}{\partial\beta_1} & \frac{\partial\hon}{\partial\beta_2}\end{pmatrix}^T.
 $$
Then  we can express $-\dpon=\frac{1}{2N}\sumn h_{\theta,n}h_{\theta,n}^T$. 
Suppose that $-\dpon$ is not a positive definite matrix. 
This implies the existence of $\lambda\ne 0$ which satisfies $\frac{1}{2N}\sumn \lambda^T h_{\theta,n} h_{\theta,n}^T \lambda=0$, implying that $h_{\theta,n}^T \lambda=0$ for all $n=1,...,N$. 
Define 
\begin{align*}
&J = \left(h_{\theta,1} \ h_{\theta,2}\ \cdots\ h_{\theta,n}\right)\\
&\quad=\begin{pmatrix}
    \frac{{\partial h_1(\to)}}{{\partial \omega_1}} & \cdots & \cdots &\frac{{\partial\aw}}{{\partial\omega_1}}+\frac{{\partial\agij}}{{\partial\omega_1}}h_{n-1}(\to)+\agij\frac{{\partial h_{n-1}(\to)}}{{\partial\omega_1}}+\frac{{\partial\ab}}{{\partial\omega_1}} Z_{n-1}^2
    \\\frac{{\partial h_1(\to)}}{{\partial \omega_2}} & \cdots & \cdots &\frac{{\partial\aw}}{{\partial\omega_2}}+\frac{{\partial\agij}}{{\partial\omega_2}}h_{n-1}(\to)+\agij\frac{{\partial h_{n-1}(\to)}}{{\partial\omega_2}}+\frac{{\partial\ab}}{{\partial\omega_2}} Z_{n-1}^2
    \\\frac{{\partial h_1(\to)}}{{\partial \gamma_1}} & \cdots & \cdots &\frac{{\partial\aw}}{{\partial\gamma_1}}+\frac{{\partial\agij}}{{\partial\gamma_1}}h_{n-1}(\to)+\agij\frac{{\partial h_{n-1}(\to)}}{{\partial\gamma_1}}+\frac{{\partial\ab}}{{\partial\gamma_1}} Z_{n-1}^2
    \\\frac{{\partial h_1(\to)}}{{\partial \gamma_2}} & \cdots & \cdots &\frac{{\partial\aw}}{{\partial\gamma_2}}+\frac{{\partial\agij}}{{\partial\gamma_2}}h_{n-1}(\to)+\agij\frac{{\partial h_{n-1}(\to)}}{{\partial\gamma_2}}+\frac{{\partial\ab}}{{\partial\gamma_2}} Z_{n-1}^2
    \\\frac{{\partial h_1(\to)}}{{\partial \beta_1}} & \cdots & \cdots &\frac{{\partial\aw}}{{\partial\beta_1}}+\frac{{\partial\agij}}{{\partial\beta_1}}h_{n-1}(\to)+\agij\frac{{\partial h_{n-1}(\to)}}{{\partial\beta_1}}+\frac{{\partial\ab}}{{\partial\beta_1}} Z_{n-1}^2
    \\\frac{{\partial h_1(\to)}}{{\partial \beta_2}} & \cdots & \cdots &\frac{{\partial\aw}}{{\partial\beta_2}}+\frac{{\partial\agij}}{{\partial\beta_2}}h_{n-1}(\to)+\agij\frac{{\partial h_{n-1}(\to)}}{{\partial\beta_2}}+\frac{{\partial\ab}}{{\partial\beta_2}} Z_{n-1}^2
\end{pmatrix},
\end{align*}
where
\begin{align*}
    &\frac{{\partial\aw}}{{\partial\theta_k}}=\yaya\frac{{\partial\wah}}{{\partial\theta_k}}+\yayb\frac{{\partial\wbh}}{{\partial\theta_k}}+\ybya\frac{{\partial\wch}}{{\partial\theta_k}}+\ybyb\frac{{\partial\wdh}}{{\partial\theta_k}},\\
    &\frac{{\partial\agij}}{{\partial\theta_k}}=\yaya\frac{{\partial\gah}}{{\partial\theta_k}}+\yayb\frac{{\partial\gbh}}{{\partial\theta_k}}+\ybya\frac{{\partial\gch}}{{\partial\theta_k}}+\ybyb\frac{{\partial\gdh}}{{\partial\theta_k}},\\
    &\frac{{\partial\ab}}{{\partial\theta_k}}=\yaya\frac{{\partial\bah}}{{\partial\theta_k}}+\yayb\frac{{\partial\bbh}}{{\partial\theta_k}}+\ybya\frac{{\partial\bch}}{{\partial\theta_k}}+\ybyb\frac{{\partial\bdh}}{{\partial\theta_k}},
\end{align*}
for any $k\in\{1,2,3,4,5,6\}$ and $\theta=(\theta_1,\theta_2,\theta_3,\theta_4,\theta_5,\theta_6)=(\wa,\wb,\ga,\gb,\ba,\bb)$. 
Since $Z_i$'s are nondegenerated, $J$ is of full rank almost surely. 
Therefore, $J^T\lambda=0$ a.s. implies $\lambda = 0$ a.s., which is contradiction. 
\endproof

\proof{Proof of Theorem \ref{thm3}.}
By mean-value theorem, there exists $\theta^*$ between $\hat{\theta}$ and $\to$ such that\\ $\hat{S}_{N,M}(\hat{\theta})-\phonm=-\phonm=\triangledown\hat{S}_{N,M}(\ts)(\hat{\theta}-\to)$.
If $-\triangledown\hat{S}_{N,M}(\theta^*)\xrightarrow{\ p\ }-\dpon$, which is a positive definite matrix, convergence rates of $|\phonm|$ and $|\hat{\theta}-\to|$ are equivalent. 
Therefore, proof of Theorem \ref{thm3} is equivalent to show (a) $\phonm=O_p(M^{-\frac{1}{4}}+N^{-\frac{1}{2}})$ and (b) $\maxnorm{\triangledown\hat{S}_{N,M}(\ts)-\dpon}=o_p(1)$.
Consider (a). 
By Assumption Lemma \ref{lem bounded}(d) and Assumption \ref{ass2}(b), we have
\begin{align*}
    \lonenorm{\phonm-\phon}&=\frac{1}{2N}\lonenorm{\sumn\frac{{\partial\hon}}{{\partial\theta}}
    \frac{1}{\hon^2}(RV_n-\isign)}\\
    &\le\frac{1}{2N}\sumn\ltwonorm{\frac{{\partial\hon}}{{\partial\theta}}
    \frac{1}{\hon^2}}\ltwonorm{RV_n-\isign}\\
    &\le CM^{-\frac{1}{4}}.
\end{align*}
Then we have
\begin{align*}
    \phonm-\pon=\phonm&=-\frac{1}{2N}\sumn\frac{{\partial\hon}}{{\partial\theta}}
    \frac{1}{\hon}\frac{(RV_n-\isign)+\xi_n}{\hon}\\
    &=-\frac{1}{2N}\sumn\frac{{\partial\hon}}{{\partial\theta}}
    \frac{1}{\hon}\frac{\xi_n}{\hon}+O_p(M^{-\frac{1}{4}}).
\end{align*}
By It\^o's lemma and It\^o's isometry, for $j\in\{1,2,3,4,5,6\}$, we have
\begin{align*}
    &E\left[\left(\frac{1}{2N}\sumn\frac{{\partial\hon}}{{\partial\theta_j}}
    \frac{1}{\hon}\frac{\xi_n}{\hon}\right)^2\right]\\
    &=\frac{1}{4N^2}E\left[\sumn\left(\frac{{\partial\hon}}{{\partial\theta_j}}\right)^2
    \left(\frac{1}{\hon}\right)^2\left(\frac{\xi_n}{\hon}\right)^2\right]\\
    &=\frac{1}{4N^2}E\left[\sumn\left(\frac{{\partial\hon}}{{\partial\theta_j}}\right)^2
    \left(\frac{1}{\hon}\right)^2\frac{E[\xi_n^2\sfilta]}{\hon^2}\right]\\
    &\le C\frac{1}{N^2}E\left[\sumn\left(\frac{{\partial\hon}}{{\partial\theta_j}}\right)^2
    \left(\frac{1}{\hon}\right)^2\frac{E[Z_n^4\sfilta]}{\hon^2}\right]\\
    &\le C\frac{1}{N^2}\sumn\ltwonorm{\left(\frac{{\partial\hon}}{{\partial\theta_j}}\right)^2
    \left(\frac{1}{\hon}\right)^2}\ltwonorm{\frac{E[Z_n^4\sfilta]}{\hon^2}}\\
    &=O_p(N^{-1}),
\end{align*}
where the last equality is hold by Lemma \ref{lem bounded}(d) and Assumption \ref{ass1}(c). 
Therefore, the statement of (a) is proved.

Consider (b). 
By the triangular inequality, we have
\begin{align*}
    \maxnorm{\triangledown\hat{S}_{N,M}(\ts)-\dpon}
    &\le\maxnorm{\triangledown\hat{S}_{N,M}(\ts)-\dphonm}\tag{b-1}\\
    &\quad+\maxnorm{\dphonm-\dpon}\tag{b-2}.
\end{align*}
For (b-1),  let $U_n=max_{j,k,l\in\{1,2,...,6\}^3} \supt\abs{\frac{\partial^3\lhn}{\partial\theta_j\partial\theta_k\partial\theta_l}}$. 
By Taylor expansion and mean value theorem, following inequality is satisfied for $\theta^{**}$ between $\to$ and $\ts$:
\begin{align*}
    \maxnorm{\triangledown\hat{S}_{N,M}(\ts)-\dphonm}
    &\le\frac{1}{2N}\sumn\maxnorm{\frac{\partial^3\hat{q}_{N,M}(\theta^{**})}{\partial\theta_j\partial\theta_k\partial\theta_l}}\maxnorm{\ts-\to}\\
    &\le C\frac{1}{2N}\sumn U_n\maxnorm{\ts-\to}=o_p(1),
\end{align*}
where the last line is due to Theorem \ref{thm2} and Lemma \ref{lem4}. 
For (b-2), by Lemma \ref{lem bounded}(d) and Assumption \ref{ass2}(b), we have
\begin{align*}
    &\maxnorm{\dphonm-\dphon}\\
    &=\maxnorm{\frac{1}{2N}\sumn\left[\frac{2}{\hn^3}\dhot\dhott-\frac{1}{\hn^2}\ddhott\right]\left(RV_n-\isign\right)}\\
    &=O_p(M^{-\frac{1}{4}}).
\end{align*}
Note that we have
\begin{align*}
    \quad\eta_n&=\triangledown\hat{S}_{N}(\to)-\dpon\\
    &=\frac{1}{2N}\sumn\ddhott\hon^{-1}\left(\frac{\isign-\hon}{\hon}\right)\\
    &\quad-2\dhot\dhott\hon^{-2}\frac{\isign-\hon}{\hon}\\
    &=\frac{1}{2N}\sumn\frac{\xi_n}{\hon}\left[\ddhott\hon^{-1}-2\dhot\dhott\hon^{-2}\right].
\end{align*}
We can easily show that $\maxnorm{\eta_n^2}\le C\frac{1}{N^2}\sumn\maxnorm{\frac{E[Z_n^4\sfilta]}{\hon^2}}=O_p(N^{-1})$. As a result, we have
\begin{align*}
    \dphonm&=\dphon+O_p(M^{-\frac{1}{4}})\\
    &=\dpon+\eta_n+O_p(M^{-\frac{1}{4}})\\
    &=\dpon+O_p(N^{-\frac{1}{2}})+O_p(M^{-\frac{1}{4}}). 
\end{align*}
\endproof

\subsubsection{Proof of Theorem \ref{thm4}}
\proof{Proof of Theorem \ref{thm4}.}
For any $\lambda\in\mathbb{R}^6$, let $v_n=\dhot\hon^{-2}\xi_n$ and $\kappa_n = \lambda^T v_n$. Since $\kappa_n$ is martingale difference, $E(\kappa_n^2)<\infty$. 
Also, $\kappa_n$ is stationary and ergodic by Assumption \ref{ass1}(d). 
Let $\frac{1}{\sqrt{N}}\sumn \kappa_n^2 \xrightarrow{\ p\ } \kappa$.
By martingale CLT, $\frac{1}{\sqrt{N}}\kappa^{-\frac{1}{2}}\sumn \kappa_n\xrightarrow{\ d\ }N(0,1)$.
Let $\frac{1}{4N}\sumn v_n v_n^T \xrightarrow{\ p\ } V$.
Then, by Cramer-Wold device, we have
\begin{align*}
-\sqrt{N}V^{-\frac{1}{2}}\phon=\frac{\sqrt{N}}{2N}V^{-\frac{1}{2}}\sumn v_n \xrightarrow{\ d\ }N(0,I_6),
\end{align*}
where $I_k$ denotes k by k identity matrix.
Furthermore, let
 $$-\dpon=\frac{1}{2N}\sumn\left[\dhot\dhott\hon^{-2} \right] \xrightarrow{\ p\ }W$$ for positive definite matrix $W$. 
By mean-value theorem, there exists $\ts$ between $\th$ and $\to$ which satisfies
\begin{align*}
    \hat{S}_{N,M}(\hat{\theta})-\phonm&=-\phonm=\triangledown\hat{S}_{N,M}(\ts)(\hat{\theta}-\to).
\end{align*}
Then  we have
\begin{align}
\begin{split}
    \sqrt{N}(\hat{\theta}-\to)&=-\sqrt{N}\triangledown\hat{S}_{N,M}(\ts)^{-1}\phonm\\
    &=\sqrt{N}(W^{-1}+o_p(1))(\phon+O_p(M^{-\frac{1}{4}}))\\
    &=\sqrt{N}\phon W^{-1}+O_p(N^{\frac{1}{2}}M^{-\frac{1}{4}})+o_p(1).
\end{split}\label{eq normality mvt}
\end{align}
Thus, we can conclude that
\begin{align*}
    (W^{-1}VW^{-1})^{-\frac{1}{2}}\sqrt{N}(\hat{\theta}-\to)\xrightarrow{\ d\ } N\left(0,I_6\right). 
\end{align*}
\endproof

\subsubsection{Proof of Proposition \ref{prop2}}
\proof{Proof of Proposition \ref{prop2}.}
We first consider $\hat{V}$.
Let 
\begin{eqnarray*}
&&I(\theta) = \frac{1}{4N}\sumn \left[\dht\dhtt\Big|_{\theta = \theta}\ h_n(\theta)^{-4} (RV_n -\ h_n(\theta)) ^2\right] = \frac{1}{4N}\sumn \iota_n(\theta), \cr
&&\tilde{I}(\theta) = \frac{1}{4N}\sumn \left[\dht\dhtt\Big|_{\theta = \theta}\ h_n(\theta)^{-4} \xi_n ^2 \right] = \frac{1}{4N}\sumn \tilde{\iota}_n(\theta).
\end{eqnarray*}
Then, $\hat{V} = I(\th)$ and we have
\begin{equation}\label{eq vconrgence 1}
	\maxnorm{I(\th) - V} \le \maxnorm{I(\th) - I(\to)} +  \Big\Vert I(\to) - V \Big\Vert_{max}. 
\end{equation}

First, we show that the convergence of $I(\th)$ to $ I(\to)$ is equivalent to that of $\th$ to $\to$.
Similar to the proof of Lemma \ref{lem4}(a), under Lemma \ref{lem bounded}(d) and Assumption \ref{ass2}(c), we can show  $U^{'}_n = \max_{j} \supB \left|\frac{\partial{\iota_n}(\theta)}{\partial{\theta_j}}\right| = O_p(1)$ for $j \in \{1,2,3,4,5,6\}$. 
For large $N$ and $M$, by mean value theorem and Taylor expansion, we have
\begin{align*}
    \maxnorm{I(\th)-\I(\to)}&\le \frac{C}{4N}\sumn U^{'}_n \maxnorm{\th-\to}.
\end{align*}
Since Theorem \ref{thm3} shows that $\th$ converges to  $\theta_0 $ with the convergence rate $N^{-1/2} +M^{-1/4}$, we have
\begin{eqnarray*}
	\maxnorm{I(\th)-\I(\to)}  = O_p (N^{-1/2} +M^{-1/4}).
\end{eqnarray*}

For  the second term of right hand side of Equation \eqref{eq vconrgence 1}, we have
\begin{align}\label{eq consistency of V}
\Big\Vert I(\to) - V \Big\Vert_{max} \le \maxnorm{I(\to) - \tilde{I}(\to)}+\maxnorm{\tilde{I}(\to)-V}.
\end{align}
We consider the first term of right hand side of Equation \eqref{eq consistency of V}.
For $j \in \{1,2,3,4,5,6\}$, we have
\begin{align*}
   &\frac{1}{4N}\sumn E\left[\frac{1}{4N}\sumn \left(\frac{\partial{\hon}}{\partial{\theta_j}}\right)^2\ \hon^{-4} \{\left(RV-\hon\right)^2 - \xi ^2\}\right] \\
    \ &\le \frac{C}{4N}\sumn \ltwonorm{\hon^{-1} \left\{(RV-\hon)^2 - \left(\isign -\hon\right)^2\right\}}\\
    \ &\le \frac{C}{4N}\sumn \lfournorm{RV-\isign}\lfournorm{\frac{RV-\isign+2\xi}{\hon}}\\
    \ &\le CM^{-\frac{1}{2}},
\end{align*} 
 where the first inequality is hold by Lemma \ref{lem bounded}(d) and the last inequality is hold by Assumption \ref{ass3}. 
Consequently, we have $\maxnorm{I(\to) - \tilde{I}(\to)} = O_p(M^{-\frac{1}{2}})$ and $\maxnorm{\tilde{I}(\to)-V} = o_p(1)$.
In conclusion, we have
$$
\maxnorm{I(\th)-V} = o_p(1) + O_p (N^{-1/2} +M^{-1/4}).
$$
Similarly, we can show that $\maxnorm{\hat{W}-W} = o_p(1) + O_p (N^{-1/2} +M^{-1/4})$ as well. 

\endproof

\subsubsection{Proof of Theorem \ref{thm5}} \label{proof wald}
\proof{Proof of Theorem \ref{thm5}.} By multiplying both sides of Equation \eqref{eq normality mvt} by $R$, we obtain
\begin{align*}
    \sqrt{N}R(\th-\to) = -R\sqrt{N}\phon W^{-1} + O_p(N^{\frac{1}{2}}M^{-\frac{1}{4}})+o_p(1).
\end{align*}
By Theorem \ref{thm4}, we have
\begin{align*}
    [RW^{-1}VW^{-1}R^T]^{\frac{1}{2}}\sqrt{N}(R\th-r)\xrightarrow{\ d\ }N(0,I_v).
\end{align*}
Then, by continuous mapping theorem, we have
\begin{align*}
    N(R\th-r)^T(RW^{-1}VW^{-1}R^T)^{-1}(R\th-r)\xrightarrow{\ d\ }\chi^2(v).
\end{align*}
% We can show that the estimates of $W$ and $V$ are consistent estimators as follows:
% \begin{align*}
%     &\frac{1}{4N}\sumn\left[\dht\dhtt\Big|_{\theta = \th}\ {\hat{h}_n(\theta)}^{-4} \hat{D}_n^2\right] = \hat{W} \xrightarrow{\ p\ }W,\\
%     &\frac{1}{2N}\sumn\left[\dht\dhtt\Big|_{\theta = \th}\ {\hat{h}_n(\theta)}^{-2} \right] = \hat{V} \xrightarrow{\ p\ }V. 
% \end{align*}
In Proposition \ref{prop2}, we already showed that $\hat{V} \xrightarrow{\ p\ }V$ and $\hat{W} \xrightarrow{\ p\ }W$.
Consequently, we have
\begin{align*}
    T_{N,M}= N(R\th-r)^T(R\hat{W}^{-1}\hat{V}\hat{W}^{-1}R^T)^{-1}(R\th-r)\xrightarrow{\ d\ }\chi^2(v). 
\end{align*}
\endproof

\clearpage

% Table generated by Excel2LaTeX from sheet 'size alpha test'
\begin{table}[htbp]
  \centering
  \caption{Size $\alpha$ test results for the Wald-type statistic}
    \begin{tabular}{llccccccccc}
    \toprule
          &       & \multicolumn{4}{c}{Under $H_0$} &       & \multicolumn{4}{c}{Under $H_a$} \\
\cmidrule{3-11}    $N$   & $M$   & 0.1   & 0.05  & 0.025 & 0.01  &       & 0.1   & 0.05  & 0.025 & 0.01 \\
    \midrule
    250   & 390   & 0.195 & 0.121 & 0.041 & 0.012 &       & 0.634 & 0.527 & 0.314 & 0.138 \\
          & 2340  & 0.162 & 0.100 & 0.032 & 0.008 &       & 0.887 & 0.835 & 0.720 & 0.475 \\
          & 4680  & 0.154 & 0.095 & 0.029 & 0.008 &       & 0.951 & 0.920 & 0.832 & 0.664 \\
          & 23400 & 0.158 & 0.086 & 0.029 & 0.007 &       & 0.994 & 0.992 & 0.977 & 0.921 \\
          &       &       &       &       &       &       &       &       &       &  \\
    500   & 390   & 0.146 & 0.092 & 0.035 & 0.008 &       & 0.880 & 0.826 & 0.628 & 0.354 \\
          & 2340  & 0.115 & 0.063 & 0.016 & 0.002 &       & 0.991 & 0.984 & 0.963 & 0.882 \\
          & 4680  & 0.128 & 0.060 & 0.017 & 0.002 &       & 1.000 & 0.999 & 0.988 & 0.966 \\
          & 23400 & 0.132 & 0.070 & 0.019 & 0.002 &       & 1.000 & 1.000 & 1.000 & 0.999 \\
          &       &       &       &       &       &       &       &       &       &  \\
    750   & 390   & 0.121 & 0.066 & 0.021 & 0.004 &       & 0.962 & 0.932 & 0.831 & 0.601 \\
          & 2340  & 0.110 & 0.063 & 0.017 & 0.002 &       & 1.000 & 1.000 & 0.997 & 0.982 \\
          & 4680  & 0.111 & 0.065 & 0.019 & 0.002 &       & 1.000 & 1.000 & 1.000 & 0.998 \\
          & 23400 & 0.113 & 0.063 & 0.016 & 0.005 &       & 1.000 & 1.000 & 1.000 & 1.000 \\
          &       &       &       &       &       &       &       &       &       &  \\
    1000  & 390   & 0.135 & 0.076 & 0.023 & 0.003 &       & 0.988 & 0.981 & 0.930 & 0.793 \\
          & 2340  & 0.109 & 0.066 & 0.013 & 0.001 &       & 1.000 & 1.000 & 1.000 & 0.998 \\
          & 4680  & 0.104 & 0.063 & 0.016 & 0.000 &       & 1.000 & 1.000 & 1.000 & 1.000 \\
          & 23400 & 0.106 & 0.055 & 0.013 & 0.001 &       & 1.000 & 1.000 & 1.000 & 1.000 \\
    \bottomrule
    \end{tabular}%
    \label{simulation table1}
    \\~\\
\begin{flushleft}
\small{\textit{Notes.} This table presents Wald-type test rejection rates under the null and alternative hypothesis for $\alpha=0.1,0.05,0.025,0.01$, $N = 250, 500, 750, 1000$, and $M = 390, 2340, 4680, 23400$.}
\end{flushleft}
\end{table}%

\clearpage

% Table generated by Excel2LaTeX from sheet 'empirical result'
\begin{table}[htbp]
  \centering
  \caption{SG-It\^o model parameter estimation and hypothesis test results based on the realized volatility estimates}
    \begin{tabular}{cccccccc}
    \toprule
          & \multicolumn{7}{c}{Models} \\
\cmidrule{2-8}    Parameters & (\romannum{1})     & (\romannum{2})     & (\romannum{3})     & (\romannum{4})     & (\romannum{5})     & (\romannum{6})     & (\romannum{7}) \\
    \midrule
    $\wa$ & 0.024$^{***}$ & 0.015$^{***}$ & 0.020$^{***}$ & 0.025$^{***}$ & 0.024$^{***}$ & 0.017$^{***}$ & 0.027$^{***}$ \\
          & (0.004) & (0.005) & (0.004) & (0.007) & (0.007) & (0.006) & (0.004) \\
    $\ga$ & 0.671$^{***}$ & 0.679$^{***}$ & 0.686$^{***}$ & 0.796$^{***}$ & 0.743$^{***}$ & 0.734$^{***}$ & 0.684$^{***}$ \\
          & (0.037) & (0.037) & (0.034) & (0.045) & (0.051) & (0.050) & (0.038) \\
    $\ba$ & 0.130$^{***}$ & 0.124$^{***}$ & 0.149$^{***}$ & 0.154$^{***}$ & 0.154$^{***}$ & 0.126$^{***}$ & 0.167$^{***}$ \\
          & (0.024) & (0.019) & (0.021) & (0.020) & (0.026) & (0.031) & (0.028) \\
    $\wb$ & 0.053$^{***}$ & 0.039$^{***}$ & 0.044$^{***}$ & 0.010 & 0.010 & 0.022$^{**}$ & 0.012 \\
          & (0.014) & (0.008) & (0.012) & (0.020) & (0.019) & (0.011) & (0.009) \\
    $\gb$ & 0.814$^{***}$ & 0.738$^{***}$ & 0.820$^{***}$ & 0.522$^{***}$ & 0.702$^{***}$ & 0.799$^{***}$ & 0.830$^{***}$ \\
          & (0.072) & (0.048) & (0.059) & (0.078) & (0.084) & (0.071) & (0.050) \\
    $\bb$ & 0.136$^{***}$ & 0.212$^{***}$ & 0.130$^{***}$ & 0.167$^{***}$ & 0.160$^{***}$ & 0.143$^{***}$ & 0.120$^{***}$ \\
          & (0.030) & (0.032) & (0.033) & (0.042) & (0.030) & (0.036) & (0.021) \\
          &       &       &       &       &       &       &  \\
    Wald  & 30.252$^{***}$ & 52.567$^{***}$ & 41.563$^{***}$ & 32.122$^{***}$ & 2.573 & 35.018$^{***}$ & 10.104$^{**}$ \\
    Statistic & (0.000)  & (0.000)  & (0.000)  & (0.000)  & (0.462)  & (0.000)  & (0.018)  \\
    \bottomrule
    \end{tabular}%
  \label{parameter table}%
 \\~\\
\begin{flushleft}
\small{\textit{Notes.} This table represents SG-It\^o model parameter estimation and hypothesis test results based on the realized volatility estimates. Models (\romannum{1})--(\romannum{7}) are constructed to examine the following effects on the volatility process: (\romannum{1}) leverage (previous-day market return), (\romannum{2}) leverage (overnight return), (\romannum{3}) Chinese stock market movement, (\romannum{4}) pre-holiday, (\romannum{5}) post-holiday, (\romannum{6}) abnormal trading volume, and (\romannum{7}) aggregate liquidity. The Wald-type statistics are from the Wald-type test under the null hypothesis $H_0: \{\wa = \wb,\ \ga = \gb,\ \ba = \bb\}$. For the parameter estimation, intraday S\&P 500 index data spanning from January 1, 2015, to December 31, 2018, are used. The numbers in parentheses under parameter estimates and Wald-type statistics indicate standard errors and $p$-values, respectively. $^{***}$ and $^{**}$ on coefficients and Wald-type statistics denote statistical significance at the 1\% and 5\% level, respectively.}
\end{flushleft}
\end{table}%

\clearpage

% Table generated by Excel2LaTeX from sheet 'empirical result'
\begin{table}[htbp]
  \centering
  \caption{Integrated form of SG-It\^o model parameter estimates}
    \begin{tabular}{cccccccc}
    \toprule
          & \multicolumn{7}{c}{Models} \\
\cmidrule{2-8}    Parameters & (\romannum{1})     & (\romannum{2})     & (\romannum{3})     & (\romannum{4})     & (\romannum{5})     & (\romannum{6})     & (\romannum{7}) \\
    \midrule
    $\wah$ & 0.026 & 0.016 & 0.022 & 0.027 & 0.026 & 0.018 & 0.029 \\
    $\gah$ & 0.671 & 0.679 & 0.686 & 0.796 & 0.743 & 0.734 & 0.684 \\
    $\bah$ & 0.117 & 0.111 & 0.136 & 0.150 & 0.146 & 0.117 & 0.153 \\
          &       &       &       &       &       &       &  \\
    $\wbh$ & 0.039 & 0.029 & 0.033 & 0.020 & 0.019 & 0.021 & 0.020 \\
    $\gbh$ & 0.749 & 0.681 & 0.769 & 0.608 & 0.716 & 0.764 & 0.785 \\
    $\bbh$ & 0.122 & 0.190 & 0.119 & 0.162 & 0.151 & 0.133 & 0.110 \\
          &       &       &       &       &       &       &  \\
    $\wch$ & 0.042 & 0.030 & 0.035 & 0.017 & 0.019 & 0.022 & 0.022 \\
    $\gch$ & 0.729 & 0.737 & 0.732 & 0.683 & 0.728 & 0.768 & 0.723 \\
    $\bch$ & 0.127 & 0.121 & 0.145 & 0.129 & 0.143 & 0.122 & 0.162 \\
          &       &       &       &       &       &       &  \\
    $\wdh$ & 0.057 & 0.044 & 0.046 & 0.011 & 0.011 & 0.024 & 0.012 \\
    $\gdh$ & 0.814 & 0.738 & 0.820 & 0.522 & 0.702 & 0.799 & 0.830 \\
    $\bdh$ & 0.133 & 0.206 & 0.127 & 0.139 & 0.148 & 0.139 & 0.117 \\
    \bottomrule
    \end{tabular}%
  \label{gparameter table}%
 \\~\\
\begin{flushleft}
\small{\textit{Notes.} This table presents the integrated form of SG-It\^o model parameter estimates (i.e., $\hat{\theta^h}$) suggested in Theorem \ref{thm1}(b). 
Models (\romannum{1})--(\romannum{7}) are constructed to examine the following effects on the volatility process: (\romannum{1}) leverage (previous-day market return), (\romannum{2}) leverage (overnight return), (\romannum{3}) Chinese stock market movement, (\romannum{4}) pre-holiday, (\romannum{5}) post-holiday, (\romannum{6}) abnormal trading volume, and (\romannum{7}) aggregate liquidity.}
\end{flushleft}
\end{table}

\clearpage

% Table generated by Excel2LaTeX from sheet 'benchmark params'
\begin{table}[htbp]
  \centering
  \caption{Estimation of the GARCH, GARCH-It\^o, and RS-GARCH model parameters}
    \begin{tabular}{cccccccccc}
    \toprule
          &       &       & \multicolumn{7}{c}{RS-GARCH} \\
\cmidrule{4-10}    Parameters & GARCH & GARCH-It\^o & (\romannum{1})     & (\romannum{2})     & (\romannum{3})     & (\romannum{4})     & (\romannum{5})     & (\romannum{6})     & (\romannum{7}) \\
    \midrule
    $\omega^L_1$ & 0.023 & 0.021 & 0.016 & 0.001 & 0.017 & 0.010 & 0.029 & 0.009 & 0.027 \\
    $\gamma^L_1$ & 0.772 & 0.738 & 0.805 & 0.764 & 0.805 & 0.835 & 0.787 & 0.696 & 0.749 \\
    $\beta^L_1$ & 0.204 & 0.153 & 0.043 & 0.159 & 0.070 & 0.165 & 0.178 & 0.192 & 0.212 \\
          &       &       &       &       &       &       &       &       &  \\
    $\omega^L_2$ &       &       & 0.086 & 0.054 & 0.070 & 0.052 & 0.000 & 0.073 & 0.012 \\
    $\gamma^L_2$ &       &       & 0.853 & 0.800 & 0.723 & 0.613 & 0.725 & 0.767 & 0.841 \\
    $\beta^L_2$ &       &       & 0.147 & 0.200 & 0.277 & 0.330 & 0.275 & 0.233 & 0.159 \\
    \bottomrule
    \end{tabular}%
  \label{all gparameter table}%
\\~\\
\begin{flushleft}
\small{\textit{Notes.} This table presents paremeter estimates of the GARCH, GARCH-It\^o, and RS-GARCH models.
The RS-GARCH models models (\romannum{1})--(\romannum{7}) are constructed to examine the following effects on the volatility process: (\romannum{1}) leverage (previous-day market return), (\romannum{2}) leverage (overnight return), (\romannum{3}) Chinese stock market movement, (\romannum{4}) pre-holiday, (\romannum{5}) post-holiday, (\romannum{6}) abnormal trading volume, and (\romannum{7}) aggregate liquidity.}
\end{flushleft}
\end{table}

\clearpage

% Table generated by Excel2LaTeX from sheet 'empirical result lf'
\begin{table}[htbp]
  \centering
  \caption{SG-It\^o model parameter estimation and hypothesis test results based on the low-frequency data only}
    \begin{tabular}{cccccccc}
    \toprule
          & \multicolumn{7}{c}{Models} \\
\cmidrule{2-8}    Parameters & (\romannum{1})     & (\romannum{2})     & (\romannum{3})     & (\romannum{4})     & (\romannum{5})     & (\romannum{6})     & (\romannum{7}) \\
    \midrule
    $\wa$ & 0.018$^{*}$ & 0.010 & 0.022$^{*}$ & 0.031 & 0.031 & 0.010 & 0.021$^{**}$ \\
          & (0.010) & (0.017) & (0.012) & (0.024) & (0.023) & (0.012) & (0.010) \\
    $\ga$ & 0.744$^{***}$ & 0.758$^{***}$ & 0.705$^{***}$ & 0.777$^{***}$ & 0.736$^{***}$ & 0.760$^{***}$ & 0.707$^{***}$ \\
          & (0.098) & (0.085) & (0.062) & (0.067) & (0.086) & (0.110) & (0.052) \\
    $\ba$ & 0.123 & 0.158$^{**}$ & 0.207$^{***}$ & 0.173$^{***}$ & 0.214 & 0.162 & 0.243$^{***}$ \\
          & (0.109) & (0.073) & (0.064) & (0.045) & (0.069) & (0.099) & (0.055) \\
    $\wb$ & 0.115 & 0.059$^{*}$ & 0.082 & 0.010 & 0.010 & 0.059 & 0.033 \\
          & (0.098) & (0.035) & (0.071) & (0.078) & (0.048) & (0.072) & (0.034) \\
    $\gb$ & 0.803$^{***}$ & 0.704$^{***}$ & 0.787$^{***}$ & 0.680$^{***}$ & 0.789$^{***}$ & 0.710$^{***}$ & 0.940$^{***}$ \\
          & (0.273) & (0.125) & (0.148) & (0.138) & (0.139) & (0.202) & (0.093) \\
    $\bb$ & 0.147$^{**}$ & 0.246$^{***}$ & 0.163$^{***}$ & 0.270$^{***}$ & 0.161$^{**}$ & 0.240$^{**}$ & 0.010 \\
          & (0.069) & (0.068) & (0.046) & (0.089) & (0.064) & (0.109) & (0.048) \\
          &       &       &       &       &       &       &  \\
    Wald  & 7.139$^{*}$ & 2.723 & 5.287 & 0.983 & 0.353 & 7.603$^{*}$ & 9.443$^{**}$ \\
    Statistic & (0.068)  & (0.436)  & (0.152)  & (0.805)  & (0.950)  & (0.055)  & (0.024)  \\
    \bottomrule
    \end{tabular}%
  \label{parameter table lf}%
\\~\\
\begin{flushleft}
\small{\textit{Notes.} This table presents SG-It\^o model parameter estimation and hypothesis test results based on the low-frequency data for models (\romannum{1})--(\romannum{7}). Models (\romannum{1})--(\romannum{7}) are constructed to examine the following effects on the volatility process: (\romannum{1}) leverage (previous-day market return), (\romannum{2}) leverage (overnight return), (\romannum{3}) Chinese stock market movement, (\romannum{4}) pre-holiday, (\romannum{5}) post-holiday, (\romannum{6}) abnormal trading volume, and (\romannum{7}) aggregate liquidity.
The Wald-type statistics are from the Wald-type test under the null hypothesis $H_0: \{\wa = \wb,\ \ga = \gb,\ \ba = \bb\}$. 
For the parameter estimation, daily S\&P 500 index data spanning from January 1, 2015, to December 31, 2018 are used. 
The numbers in parentheses under parameter estimates and Wald-type statistics indicate standard error and $p$-value, respectively. $^{***}$, $^{**}$, and $^{*}$ on coefficients and Wald-type statistics denotes statistical significance at the 1\%, 5\%, and 10\% level, respectively.}
\end{flushleft}
\end{table}

\clearpage

% Table generated by Excel2LaTeX from sheet 'oos prediction'
\begin{table}[htbp]
  \centering
  \caption{Out-of-sample prediction performance of the volatility models measured by MAPEs}
    \begin{tabular}{cccccccc}
    \toprule
    & \multicolumn{7}{c}{Out-of-sample MAPEs} \\
\cmidrule{2-8} Volatility models & (\romannum{1})    & (\romannum{2})     & (\romannum{3})     & (\romannum{4})     & (\romannum{5})     & (\romannum{6})     & (\romannum{7}) \\
    \midrule
    SG-Ito & 0.560 & 0.503 & 0.552 & 0.558 & 0.567 & 0.564 & 0.572 \\
    GARCH-Ito & 0.571 & 0.571 & 0.571 & 0.571 & 0.571 & 0.571 & 0.571 \\
    RS-GARCH & 0.812 & 0.766 & 0.805 & 0.930 & 0.909 & 0.917 & 0.952 \\
    GARCH & 0.925 & 0.925 & 0.925 & 0.925 & 0.925 & 0.925 & 0.925 \\
    HAR & 0.646 & 0.646 & 0.646 & 0.646 & 0.646 & 0.646 & 0.646 \\

    \bottomrule
    \end{tabular}%
  \label{prediction table}%
\\~\\
\begin{flushleft}
\small{\textit{Notes.} This table presents out-of-sample prediction performance of the volatility models measured by MAPEs. 
Models (\romannum{1})--(\romannum{7}) are constructed to examine the following effects on the volatility process: (\romannum{1}) leverage (previous-day market return), (\romannum{2}) leverage (overnight return), (\romannum{3}) Chinese stock market movement (\romannum{4}) pre-holiday, (\romannum{5}) post-holiday, (\romannum{6}) abnormal trading volume, and (\romannum{7}) aggregate liquidity. 
Estimation window is 750 days and prediction period is from December 22, 2017, to December 31, 2018 (248 days).}
\end{flushleft}
\end{table}

\clearpage

\clearpage

\clearpage

\clearpage

\section*{Data availability statement}

The S\&P500 intraday index data is provided by the Chicago Board of Exchange (CBOE). \\
(web link: https://datashop.cboe.com/). 
Please note that the data sharing policy of CBOE restricts the redistribution of data.

\bibliography{bibliography}

\end{document}